\documentclass[aps,superscriptaddress,amsmath,amssymb,twocolumn,showpacs,floatfix,reprint]{revtex4-1}
\usepackage{bm}
\usepackage[colorlinks=true, urlcolor=blue, linkcolor=blue, citecolor=blue, pdftex]{hyperref}
\usepackage{float}
\usepackage[usenames,dvipsnames]{xcolor}
\usepackage{comment}
\usepackage[utf8]{inputenc}
\usepackage{braket}
\usepackage{graphicx}

\usepackage{algorithm}
\usepackage{algpseudocode}

\usepackage{nicefrac}

\definecolor{C0}{HTML}{1f77b4}
\definecolor{C1}{HTML}{ff7f0e}
\definecolor{C2}{HTML}{2ca02c}
\definecolor{C3}{HTML}{d62728}
\definecolor{C4}{HTML}{9467bd}
\definecolor{C5}{HTML}{8c564b}

\usepackage{lineno}

\usepackage{todonotes}

\begin{document}

\title{A simple linear algebra identity to optimize \\
Large-Scale Neural Network Quantum States}

\author{Riccardo Rende}
\thanks{These authors contributed equally.}
\affiliation{International School for Advanced Studies (SISSA), Via Bonomea 265, I-34136 Trieste, Italy}
\author{Luciano Loris Viteritti}
\thanks{These authors contributed equally.}
\affiliation{Dipartimento di Fisica, Universit\`a di Trieste, Strada Costiera 11, I-34151 Trieste, Italy}
\author{Lorenzo Bardone}
\affiliation{International School for Advanced Studies (SISSA), Via Bonomea 265, I-34136 Trieste, Italy}
\author{Federico Becca}
\affiliation{Dipartimento di Fisica, Universit\`a di Trieste, Strada Costiera 11, I-34151 Trieste, Italy}
\author{Sebastian Goldt}
\thanks{Correspondence should be addressed to rrende@sissa.it, lucianoloris.viteritti@phd.units.it, and sgoldt@sissa.it.}
\affiliation{International School for Advanced Studies (SISSA), Via Bonomea 265, I-34136 Trieste, Italy}

\date{\today}

\begin{abstract}
Neural-network architectures have been increasingly used to represent quantum many-body wave functions. These networks require a large number of variational parameters and are challenging to optimize using traditional methods, as gradient descent. Stochastic Reconfiguration (SR) has been effective with a limited number of parameters, but becomes impractical beyond a few thousand parameters. Here, we leverage a simple linear algebra identity to show that SR can be employed even in the deep learning scenario. We demonstrate the effectiveness of our method by optimizing a Deep Transformer architecture with $3 \times 10^5$ parameters, achieving state-of-the-art ground-state energy in the $J_1$-$J_2$ Heisenberg model at $J_2/J_1=0.5$ on the $10\times10$ square lattice, a challenging benchmark in highly-frustrated magnetism. This work marks a significant step forward in the scalability and efficiency of SR for Neural-Network Quantum States, making them a promising method to investigate unknown quantum phases of matter, where other methods struggle.
\end{abstract}

\maketitle

\section{Introduction}
Deep learning has become crucial in many research fields, with neural networks being the key to achieve impressive results. Well-known examples include Deep Convolutional Neural 
Networks for image recognition~\cite{imagenet2012,residuals2016} and Deep Transformers for language-related tasks~\cite{vaswani2017attention, bert2019, nlp2020}. The success of deep networks comes 
from two ingredients: architectures with a large number of parameters (often in the billions), which allow for great flexibility, and training these networks on large amounts 
of data. However, to successfully train these large models in practice, one needs to navigate the complicated and highly non-convex landscape associated with this extensive 
parameter space.

The most used methods rely on stochastic gradient descent (SGD), where the gradient of the loss function is estimated from a randomly selected subset of the training data. 
Over the years, variations of traditional SGD, such as Adam~\cite{adam2017} or AdamW~\cite{adamw2019}, have proven highly effective, leading to more accurate results. In the late 1990s, Amari and collaborators~\cite{Amari1998,Amari2018} suggested to use the knowledge of the geometric 
structure of the parameter space to adjust the gradient direction for non-convex landscapes, defining the concept of Natural Gradients. In the same years, 
Sorella~\cite{Sorella1998,Sorella2005} proposed a similar method, now known as Stochastic Reconfiguration (SR), to enhance the optimization of variational functions in quantum 
many-body systems. Importantly, this latter approach typically outperforms other methods such as SGD or Adam, leading to significantly lower variational energies. The main 
idea of SR is to exploit information about the curvature of the loss landscape, thus improving the convergence speed in landscapes which are steep in some directions and 
shallow in others~\cite{Park2020}. For physically inspired wave functions (e.g., Jastrow-Slater~\cite{Capello2005} or Gutzwiller-projected states~\cite{BeccaGutz2013}) the 
original SR formulation is a highly efficient method since there are few variational parameters, typically from~$O(10)$ to~$O(10^2)$. 

Over the past few years, neural networks have been extensively used as powerful variational \textit{Ansätze} for studying interacting spin models~\cite{Carleo2017}, and the number 
of parameters has increased significantly. Starting with simple Restricted Boltzmann Machines (RBMs)~\cite{Carleo2017,FerrariGutz2019,Nomura2017,Viteritti2022,Park2022,Nomura2023}, 
more complicated architectures as Convolutional Neural Networks (CNNs)~\cite{CNNCarleo2019,CNNXiao2018,CNNSzabo2020} and Recurrent Neural Networks 
(RNNs)~\cite{RNNCarrasquilla2020,RNNRoth2020,RNNHibatallah2022,RNNHibatallah2023} have been introduced to handle challenging systems. In particular, Deep 
CNNs~\cite{AttilaGCNN2021,LiCNN2022,HeylCNN2023,LiangCNN2022} have proven to be highly accurate for two-dimensional models, outdoing methods as Density-Matrix Renormalization 
Group (DMRG) approaches~\cite{DMRGSheng2014} and Gutzwiller-projected states~\cite{BeccaGutz2013}. These deep learning models have great performances when the number of parameters 
is large. However, a significant bottleneck arises when employing the original formulation of SR for optimization, as it is based on the inversion of a matrix of size $P\times P$, 
where $P$ denotes the number of parameters. Consequently, this approach becomes computationally infeasible as the parameter count exceeds $O(10^4)$, primarily due to the 
constraints imposed by the limited memory capacity of current-generation GPUs.
\begin{table*}[ht]
\caption{Ground-state energy on the 10$\times$10 square lattice at $J_2/J_1 = 0.5$.}
\centering
\label{table:accuracy}
\begin{tabular}{p{0.15\linewidth}p{0.25\linewidth}p{0.18\linewidth}p{0.14\linewidth}p{0.13\linewidth}p{0.05\linewidth}}
\hline\hline
\textbf{Energy per site} & \centering \textbf{Wave function} & \centering \textbf{\# parameters} &  \centering \textbf{Marshall prior}&\centering \textbf{Reference} & \textbf{Year} \\ \hline
-0.48941(1)         & \centering MLP                 & \centering 893994             &  \centering Not available  &\centering \cite{LedinNN2023}    & 2023    \\
-0.494757(12)       & \centering CNN                  & \centering Not available      &  \centering No             &\centering \cite{CNNSzabo2020}   & 2020    \\
-0.4947359(1)       & \centering Shallow CNN          & \centering 11009              &  \centering Not available  &\centering \cite{CNNXiao2018}    & 2018    \\
-0.49516(1)         & \centering Deep CNN             & \centering 7676               &  \centering Yes            &\centering \cite{CNNCarleo2019}  & 2019    \\
-0.495502(1)        & \centering PEPS + Deep CNN      & \centering 3531               &  \centering No             &\centering \cite{LiangPEPS2021}  & 2021    \\
-0.495530           & \centering DMRG                 & \centering 8192 SU(2) states  &  \centering No             &\centering \cite{DMRGSheng2014}  & 2014    \\
-0.495627(6)        & \centering aCNN                 & \centering 6538               &  \centering Yes            &\centering \cite{aCNNWang2023}   & 2023    \\
-0.49575(3)         & \centering RBM-fermionic        & \centering 2000               &  \centering Yes            &\centering \cite{FerrariGutz2019}& 2019    \\
-0.49586(4)         & \centering CNN                  & \centering 10952              &  \centering Yes            &\centering \cite{SchmittCNN2023} & 2023    \\
-0.4968(4)          & \centering RBM $(p=1)$          & \centering Not available      &  \centering Yes            &\centering \cite{ChenRBM2022}    & 2022    \\
-0.49717(1)         & \centering Deep CNN             & \centering 106529             &  \centering Yes            &\centering \cite{LiCNN2022}      & 2022    \\
-0.497437(7)        & \centering GCNN                 & \centering 67548      &  \centering No             &\centering \cite{AttilaGCNN2021} & 2023    \\
-0.497468(1)        & \centering Deep CNN             & \centering 421953             &  \centering Yes            &\centering \cite{LiangCNN2022}   & 2022    \\
-0.4975490(2)       & \centering VMC $(p=2)$          & \centering 5                  &  \centering Yes            &\centering \cite{BeccaGutz2013}  & 2013    \\
-0.497627(1)        & \centering Deep CNN             & \centering 146320             &  \centering Yes            &\centering \cite{HeylCNN2023}    & 2023    \\
-0.497629(1)        & \centering RBM+PP               & \centering 13132               &  \centering Yes            &\centering \cite{imada-prx}      & 2021    \\
\textbf{-0.497634(1)}& \centering \textbf{Deep ViT}   & \centering \textbf{267720}    &  \centering \textbf{No}    &\centering \textbf{Present work} & \textbf{2023}    \\
\hline
\end{tabular}
\end{table*}

Recently,~\citet{HeylCNN2023} made a step forward in the optimization procedure by introducing an alternative method, dubbed MinSR, to train Neural-Network Quantum States. MinSR 
does not require inverting the original $P \times P$ matrix but instead a much smaller $M \times M$ one, where $M$ is the number of configurations used to estimate the SR 
matrix. This is convenient in the deep learning setup where $P \gg M$. Most importantly, this procedure avoids allocating the $P\times P$ matrix, reducing the memory cost. 
However, this formulation is obtained by minimizing the Fubini-Study distance with an ad hoc constraint. In this work, we first use a simple 
relation from linear algebra to show, in a transparent way, that SR can be rewritten exactly in a form which involves inverting a small $M \times M$ matrix (in case of real-valued wave functions and a $2M \times 2M$ matrix for complex-valued ones) and that only a 
standard regularization of the SR matrix is required. Then, we exploit our technique to optimize a Deep Vision Transformer (Deep ViT) model, which has demonstrated exceptional 
accuracy in describing quantum spin systems in one and two spatial dimensions~\cite{ViT2023,shastry2023,Czischek2023,DiLuo2023}. Using almost~$3 \times 10^5$ variational 
parameters, we are able to achieve state-of-the-art ground-state energy in the most paradigmatic example of quantum many-body spin model, the $J_1$-$J_2$ Heisenberg model 
on square lattice:
\begin{equation}\label{eq:hamiltonian}
    \hat{H} = J_1\sum_{\langle i,j \rangle} \hat{\boldsymbol{S}}_{i}\cdot\hat{\boldsymbol{S}}_{j} + J_2\sum_{\langle\langle i,j \rangle\rangle} \hat{\boldsymbol{S}}_{i}\cdot\hat{\boldsymbol{S}}_{j} \,
\end{equation}
where $\hat{\boldsymbol{S}}_{i}=(S_i^x,S_i^y,S_i^z)$ is the $S=1/2$ spin operator at site $i$ and $J_1$ and $J_2$ are nearest- and next-nearest-neighbour antiferromagnetic couplings, 
respectively. Its ground-state properties have been the subject of many studies over the years, often with conflicting results~\cite{BeccaGutz2013,DMRGSheng2014,imada-prx}. In particular, several works focused on the highly-frustrated regime, which turns out to be challenging for numerical methods~\cite{imada-prx,HeylCNN2023,BeccaGutz2013,LiangCNN2022,AttilaGCNN2021,LiCNN2022,ChenRBM2022,SchmittCNN2023,FerrariGutz2019,aCNNWang2023,DMRGSheng2014,LiangPEPS2021,CNNCarleo2019,CNNXiao2018,CNNSzabo2020,LedinNN2023} and for this reason it is widely recognized as the benchmark model for validating new approaches. Here, we will focus on the particularly challenging case
with $J_2/J_1=0.5$ on the $10 \times 10$ cluster, where the exact solution is not known. 

Within variational methods, one of the main difficulties comes from the fact that the sign structure of the the ground state is not known for $J_2/J_1>0$. Indeed, the Marshall 
sign rule~\cite{marshall1955} gives the correct signs (for every cluster size) only when $J_2=0$. However, in order to stabilize the optimizations, many previous works imposed the Marshall sign rule as a first approximation for the sign structure (see \textit{Marshall prior} in Table~\ref{table:accuracy}). By contrast, within the present approach, we do not need to use any prior knowledge of the signs, 
thus defining a very general and flexible variational \textit{Ansatz}.  

In the following, we first show the alternative SR formulation, then discuss the Deep Transformer architecture, recently introduced by some of us~\cite{ViT2023, shastry2023} as a 
variational state, and finally present the results obtained by combining the two techniques on the $J_1$-$J_2$ Heisenberg model.

\section{Results}
\textbf{Stochastic Reconfiguration.} 
Finding the ground state of a quantum system with the variational principle involves minimizing the variational energy ${E_{\theta} = \braket{\Psi_{\theta}|\hat{H}|\Psi_{\theta}}/\braket{\Psi_{\theta}|\Psi_{\theta}}}$, 
where $\ket{\Psi_{\theta}}$ is a  variational state parametrized through a vector $\bm{\theta}$ of $P$ real parameters; in case of complex parameters, we can treat their 
real and imaginary parts separately~\cite{BeccaSorella2017}. For a system of $N$ $1/2$-spins, $\ket{\Psi_{\theta}}$ can be expanded in the computational basis 
$\ket{\Psi_{\theta}} = \sum_{\{\sigma\}} \Psi_{\theta}(\sigma)\ket{\sigma}$, where $ \Psi_{\theta}(\sigma) = \braket{\sigma|\Psi_{\theta}}$ is a map from spin configurations of 
the Hilbert space, $\{ \ket{\sigma} = \ket{\sigma_1^z, \sigma_2^z, \cdots, \sigma_N^z}, \ \sigma_i^z = \pm 1 \}$, to complex numbers. In a gradient-based optimization approach, 
the fundamental ingredient is the evaluation of the gradient of the loss, which in this case is the variational energy, with respect to the parameters $\theta_{\alpha}$ for 
$\alpha=1, \dots, P$. This gradient can be expressed as a correlation function~\cite{BeccaSorella2017}
\begin{equation}\label{eq:grad_E}
    F_{\alpha} = -\frac{\partial E_{\theta}}{\partial \theta_{\alpha}} = - 2 \Re\left[ \braket{ (\hat{H} - \braket{\hat{H}})(\hat{O}_{\alpha} - \braket{\hat{O}_{\alpha}})} \right] \ ,
\end{equation}
which involves the diagonal operator $\hat{O}_{\alpha}$ defined as ${O_{\alpha} (\sigma) = {\partial}\text{Log}[\Psi_{{\theta}}(\sigma)]/{\partial \theta_{\alpha}}}$. The 
expectation values $\langle \dots \rangle$ are computed with respect to the variational state. The SR updates~\cite{Sorella1998,BeccaSorella2017,Nomura2023} are constructed 
according to the geometric structure of the landscape:
\begin{equation}\label{eq:new_updates}
    \bm{\delta \theta} = \tau \left( S + \lambda \mathbb{I}_P \right)^{-1} \bm{F} \ ,
\end{equation}
where $\tau$ is the learning rate and $\lambda$ is a regularization parameter to ensure the invertibility of the $S$ matrix. The matrix $S$ has shape $P\times P$ and it is defined in terms of the $\hat{O}_{\alpha}$ 
operators~\cite{BeccaSorella2017}
\begin{equation}\label{eq:S_matrix}
    S_{\alpha, \beta} = \Re\left[\braket{(\hat{O}_{\alpha} - \braket{\hat{O}_{\alpha}})^{\dagger}(\hat{O}_{\beta} - \braket{\hat{O}_{\beta}})} \right]  \ .
\end{equation}

The Eq.~\eqref{eq:new_updates} defines the standard formulation of the SR, which involves the inversion of a $P\times P$ matrix, being the bottleneck of this approach when the 
number of parameters is larger than $O(10^4)$. To address this problem, we start reformulating Eq.~\eqref{eq:new_updates} in a more convenient way. For a given sample of $M$ spin 
configurations $\{\sigma_i \}$ (sampled according to $|\Psi_{\theta}(\sigma)|^2$), the stochastic estimate of $F_{\alpha}$ can be obtained as:
\begin{equation}
\label{eq:stochastic_gradient}
    \bar{F}_{\alpha} = -\Re\left[\frac{2}{M}\sum_{i=1}^M [E_{L i} - \bar{E}_L]^*[O_{\alpha i} - \bar{O}_{\alpha}] \right] \ .
\end{equation}
Here, ${E_{L i} = \braket{\sigma_i|\hat{H}|\Psi_{\theta}}/\braket{\sigma_i|\Psi_{\theta}}}$ defines the local energy for the configuration $\ket{\sigma_i}$ and $O_{\alpha i}=O_{\alpha}(\sigma_i)$; in addition, $\bar{E}_L$ and $\bar{O}_\alpha$ denote sample means. Throughout this work, we adopt the convention of using latin and greek indices to run over configurations and parameters, respectively. Equivalently, 
Eq.~\eqref{eq:S_matrix} can be stochastically estimated as
\begin{equation}
    \bar{S}_{\alpha,\beta} = \Re\left[\frac{1}{M}\sum_{i = 1}^{M}\left[O_{\alpha i} - \bar{O}_{\alpha} \right]^*\left[ O_{\beta i} - \bar{O}_{\beta} \right] \right]\ .
\end{equation}
To simplify further, we introduce ${Y_{\alpha i} = (O_{\alpha i} - \bar{O}_{\alpha})/\sqrt{M}}$ and ${\varepsilon_{i}=-2[E_{Li} - \bar{E}_L]^*/\sqrt{M}}$, allowing us to 
express Eq.~\eqref{eq:stochastic_gradient} in matrix notation as $\bm{\bar{F}} = \Re[Y \bm{\varepsilon}]$ and Eq.~\eqref{eq:S_matrix} as $\bar{S} = \Re[ YY^{\dagger}]$. Writing 
$Y = Y_R + i Y_I$ we obtain:
\begin{equation}
    \bar{S} = Y_RY_R^{T} + Y_IY_I^{T} = XX^T  \,
\end{equation}
where $X = \text{Concat}(Y_R,Y_I) \in \mathbb{R}^{P\times2M}$, the concatenation being along the last axis. Furthermore, using 
${\bm{\varepsilon} = \bm{\varepsilon}_R + i\bm{\varepsilon}_I}$, the gradient of the energy can be recast as
\begin{equation}
    \bar{\bm{F}} = Y_R\bm{\varepsilon}_R - Y_I\bm{\varepsilon}_I = X\bm{f} \ ,
\end{equation}
with $\bm{f} = \text{Concat}(\bm{\varepsilon}_R,-\bm{\varepsilon}_I) \in \mathbb{R}^{2M}$. Then, the update of the parameters in Eq.~\eqref{eq:new_updates} can be written as
\begin{equation}\label{eq:sr1}
    \bm{\delta \theta} = \tau (XX^T + \lambda \mathbb{I}_P)^{-1}X\bm{f} \ .
\end{equation}
This reformulation of the SR updates is a crucial step, which allows the use of the well-known push-through identity ${(AB + \lambda\mathbb{I}_n)^{-1}A = A(BA + \lambda\mathbb{I}_m)^{-1}}$~\cite{henderson1981,matrix-cookbook}, where $A$ and $B$ are respectively matrices with dimensions $n \times m$ and $m \times n$ (see \textit{Methods} for a derivation).
As a result, Eq.~\eqref{eq:sr1} can be rewritten as
\begin{equation}\label{eq:sr2}
    \bm{\delta \theta} = \tau X(X^TX + \lambda \mathbb{I}_{2M})^{-1}\bm{f} \ .
\end{equation}
This derivation is our first result: it shows, in a simple and transparent way, how to exactly perform the SR with the inversion of a $2M\times 2M$ matrix and, therefore, without allocating a $P\times P$ matrix. We emphasize that the last formulation is very useful in the typical deep learning setup, where $P \gg M$. Employing Eq.~\eqref{eq:sr2} instead 
of Eq.~\eqref{eq:sr1} proves to be more efficient in terms of both computational complexity and memory usage. The required operations for this new formulation are $O(M^2P) + O(M^3)$ instead of $O(P^3)$, and the memory usage is only $O(MP)$ instead of $O(P^2)$. For deep neural networks with $n_l$ layers the memory usage can be further reduced roughly to $O(MP/n_l)$ (see Ref.~\cite{novak22}). We developed a memory-efficient implementation of SR that is optimized for deployment on a multi-node GPU 
cluster, ensuring scalability and practicality for real-world applications (see \textit{Methods}). Other methods, based on iterative solvers, require $O(nMP)$ 
operations, where $n$ is the number of steps needed to solve the linear problem in Eq.~\eqref{eq:new_updates}. However, this number increases significantly for ill-conditioned matrices (the matrix $S$ has a number of zero eigenvalues equal to $P-M$), leading to many non-parallelizable iteration steps and consequently higher computational costs~\cite{netket3}.
Our proof also highlights that the diagonal-shift regularization of the $S$ matrix in parameter space [see Eq.~\eqref{eq:new_updates}] is equivalent to the same diagonal shift in sample space [see Eq.~\eqref{eq:sr2}]. Furthermore, it would be interesting to explore the applicability of regularization schemes with parameter-dependent diagonal shifts in the sample space~\cite{AttilaGCNN2021,lovato2022}. In contrast, for the MinSR update~\cite{HeylCNN2023}, a pseudo-inverse regularization is applied in order to truncate the effect of vanishing singular values during inversion.\\

\textbf{The variational wave function.} 
The architecture of the variational state employed in this work is based on the Deep Vision Transformer (Deep ViT) introduced and described in detail in Ref.~\cite{shastry2023}. In the Deep ViT, the input configuration $\sigma$ of shape 
$L\times L$ is initially split into $b\times b$ square patches, which are then linearly projected in a $d$-dimensional vector space in order to obtain an input sequence of $L^2/b^2$ vectors, i.e. $\left( \bm{x}_1, \dots, \bm{x}_{L^2/b^2} \right)$ with $\bm{x}_i \in \mathbb{R}^d$. This sequence is then processed by an encoder block employing Multi-Head Factored Attention (MHFA) mechanism~\cite{rende2023optimal,bhattacharya2020,ViT2023,rende2024queries}. This produces another output sequence of vectors $(\boldsymbol{A}_1, \dots, \boldsymbol{A}_{L^2/b^2})$ with ${\boldsymbol{A}_i \in \mathbb{R}^d}$ and can be formally implemented as follows:
\begin{equation}\label{eq:factored}
    A_{i,p}=\sum_{q=1}^d W_{p,q}\sum_{j=1}^{L^2/b^2} \alpha_{i,j}^{\mu(q)} \sum_{r=1}^dV_{q,r}x_{j,r} \ ,
\end{equation}
where $\mu(q)=\left\lceil q\ h/d\right\rceil$ select the correct attention weights of the corresponding head, being $h$ the total number of heads. The matrix $V \in \mathbb{R}^{d\times d}$ linearly transforms each input vector identically and independently. Instead, the attention matrices $\alpha^{\mu} \in \mathbb{R}^{\nicefrac{L^2}{b^2} \times \nicefrac{L^2}{b^2}}$ combine the different input vectors and the linear transformation $W \in \mathbb{R}^{d \times d}$ mixes the representations of the different heads.

Due to the global receptive field of the attention mechanism, its computational complexity scales quadratically with respect to the length of the input sequence. Note that in the standard dot product self-attention~\cite{vaswani2017attention}, the attention weights $\alpha_{i,j}$ are function of the inputs, while in the factored case employed here, the attention weights are input-independent variational parameters. This choice is the only custom modification that we employ with respect to the standard Vision Transformer architecture~\cite{dosovitskiy2021image}, which is also supported by numerical simulations and analytical arguments suggesting that queries and keys do not improve the performance in these problems~\cite{rende2024queries}. Finally, the resulting vectors $\boldsymbol{A}_i$ are processed by a two layers fully-connected network (FFN) with hidden size $2d$ and ReLu activation function. Skip connections and pre-layer normalization are implemented as described in Refs.~\cite{xiong2020layer,shastry2023}. Generally, a total of $n_l$ such blocks are stacked together to obtain a deep Transformer. This architecture operates on an input sequence, yielding another sequence of vectors $\left( \bm{y}_1, \dots, \bm{y}_{L^2/b^2} \right)$, where each $\bm{y}_i \in \mathbb{R}^d$. At the 
end, a complex-valued fully connected output layer, with ${\log[\cosh(\cdot)]}$ as activation function, is applied to $\bm{z} = \sum_i \bm{y}_i$ to produce a single number $\text{Log}[\Psi_\theta(\sigma)]$ and predict both the modulus and the phase of the input configuration (refer to Algorithm~\ref{algorithm} in \textit{Methods} for a pseudocode of the neural network architecture).

To enforce translational symmetry, we define the attention weights as $\alpha^{\mu}_{i,j} = \alpha^{\mu}_{i-j}$, thereby ensuring translational symmetry among patches. This choice reduces the computational cost during the restoration of the full translational symmetry through quantum number projection~\cite{Nomura_2021,SchmittCNN2023}. Specifically, only a summation involving $b^2$ terms is required. Under the specific assumption of translationally invariant attention weights, the factored attention mechanism can be technically implemented as a convolutional layer with $d$ input channels, $d$ output channels and a very specific choice of the convolutional kernel: $K_{p,r,k} = \sum_{q=1}^d W_{p,q} \alpha_{k}^{\mu(q)} \sum_{r=1}^dV_{q,r} \in \mathbb{R}^{d \times d \times \nicefrac{L^2}{b^2}}$. However, it is well-established that weight sharing and low-rank factorizations in learnable tensors within neural networks can lead to significantly different learning dynamics and, consequently, different final solutions~\cite{urban2017do, Ascoli2019, ingrosso2022}. Other symmetries of the Hamiltonian in Eq.~\eqref{eq:hamiltonian} [rotations, reflections ($C_{4v}$ point group) and spin parity] can also be restored within quantum number projection. As a result, the symmetrized wave function reads:
\begin{equation}\label{eq:symmViT}
    \tilde{\Psi}_{\theta}(\sigma) = \sum_{i=0}^{b^2-1}\sum_{j=0}^{7} \left[ \Psi_{\theta}({T}_i{R}_j\sigma) + \Psi_{\theta}(-T_i {R}_j\sigma)\right] \ .
\end{equation}
In the last equation, $T_i$ and $R_j$ are the translation and the rotation/reflection operators, respectively. Furthermore, due to the $SU(2)$ spin symmetry of the $J_1$-$J_2$ 
Heisenberg model, the total magnetization is also conserved and the Monte Carlo sampling (see \textit{Methods}) can be limited in the $S^z=0$ sector for the ground state search.\\

\begin{figure}[t]
\center
 \includegraphics[width=\columnwidth]{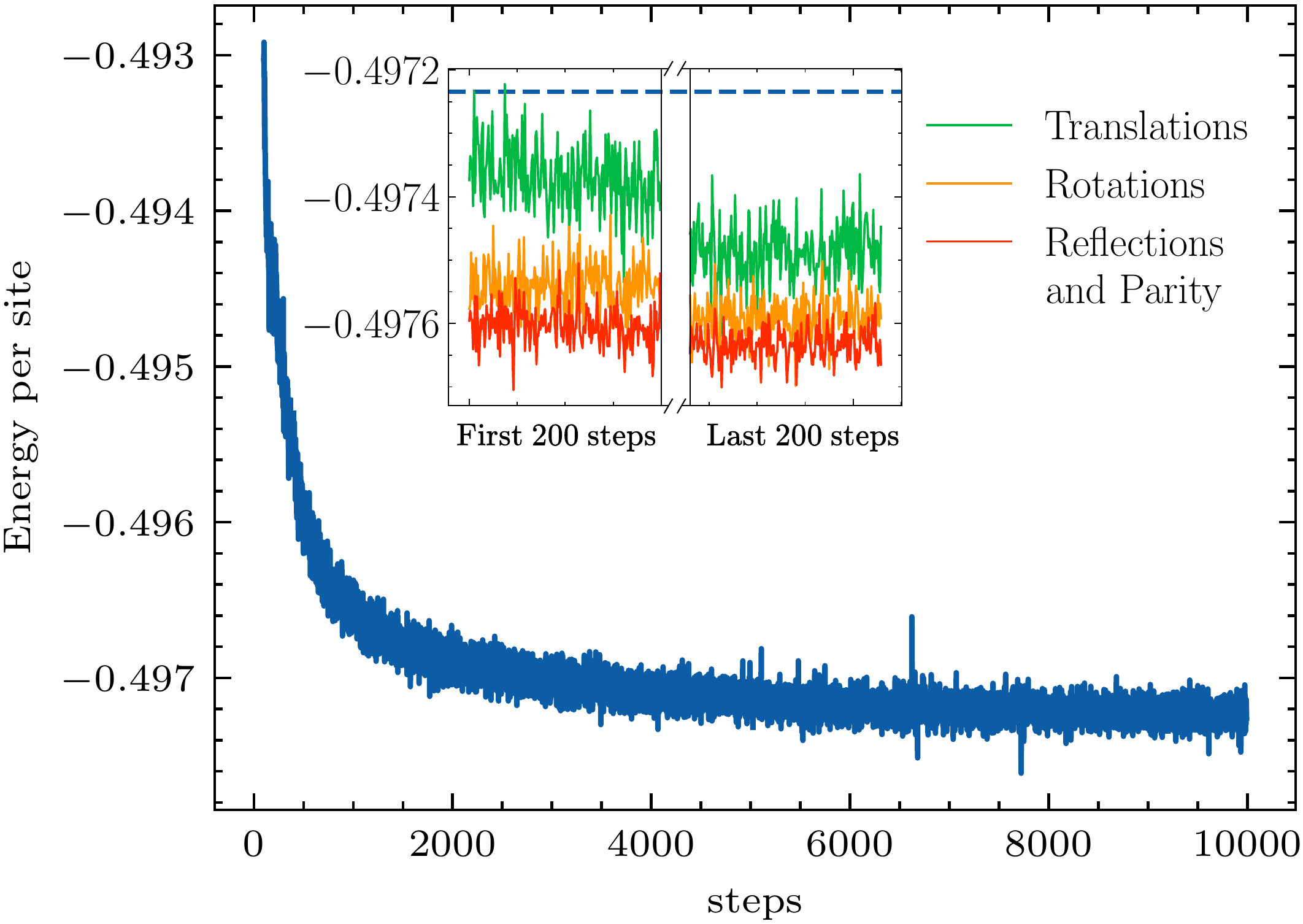}
 \caption{\label{fig:opt} \textbf{Variational energy optimization.} Optimization of the Deep ViT with patch size $b=2$, $n_l=8$ layers, embedding dimension $d=72$ and $h=12$ heads per layer, on the $J_1$-$J_2$ Heisenberg model at $J_2/J_1=0.5$ on the $10\times10$ square lattice. The first $200$ optimization steps are not shown for better readability. Inset: first and last $200$ optimization steps when recovering sequentially the full translational (green curve), rotational (orange curve) and reflections and parity (red curve) symmetries. The total number of steps after restoring the symmetries is 5000 for translations, 5000 for rotations and 4000 for reflections and parity. The mean energy obtained without quantum number projection is also reported for comparison (blue dashed line).}
\end{figure}

{\bf Numerical calculations.}
Our objective is to approximate the ground state of the $J_1$-$J_2$ Heisenberg model in the highly frustrated point $J_2/J_1=0.5$ on the $10\times 10$ square lattice. We use the 
formulation of the SR in Eq.~\eqref{eq:sr2} to optimize a variational wave function parametrized through a Deep ViT, as discussed above. The result in Table~\ref{table:accuracy} 
is achieved with the symmetrized Deep ViT in Eq.~\eqref{eq:symmViT} using $b=2$, $n_l=8$ layers, embedding dimension $d=72$, and $h=12$ heads per layer. This variational 
state has in total $267720$ real parameters (the complex-valued parameters of the output layer are treated as couples of independent real-valued parameters). Regarding the optimization protocol, we choose the learning rate $\tau=0.03$ (with cosine decay annealing) and the number of samples is 
fixed to be $M=6000$. We emphasize that using Eq.~\eqref{eq:sr1} to optimize this number of parameters would be infeasible on available GPUs: the memory requirement would be more 
than $O(10^3)$ gigabytes, one order of magnitude bigger than the highest available memory capacity. Instead, with the formulation of Eq.~\eqref{eq:sr2}, the memory requirement can be easily handled by available GPUs (see \textit{Methods}).
The simulations took four days on twenty A100 GPUs. Remarkably, as illustrated in 
Table~\ref{table:accuracy}, we are able to obtain the state-of-the-art ground-state energy using an architecture solely based on neural networks, without using any other 
regularization than the diagonal shift reported in Eq.~\eqref{eq:sr2}, fixed to $\lambda=10^{-4}$. We stress that a completely unbiased simulation, without assuming any prior 
for the sign structure, is performed, in contrast to other cases where the Marshall sign rule is used to stabilize the 
optimization~\cite{CNNCarleo2019,HeylCNN2023,imada-prx,LiCNN2022,LiangCNN2022} (see Table~\ref{table:accuracy}). 
Furthermore, we verified with numerical simulations that the final results is not affected by the Marshall prior.
This is an important point since a simple sign prior is not available 
for the majority of the models (e.g., the Heisenberg model on the triangular or Kagome lattices). We would like also to mention that the definition of a suitable architecture is
fundamental to take advantage of having a large number of parameters. Indeed, while a stable simulation with a simple regularization scheme (only defined by a finite value of
$\lambda$) is possible within the Deep ViT wave function, other architectures require more sophisticated regularizations. For example, to optimize Deep GCNNs it is necessary to 
add a temperature-dependent term to the loss function~\cite{AttilaGCNN2021} or, for Deep CNNs, a process of variance reduction and reweighting~\cite{HeylCNN2023} helps in escaping 
local minima. We also point out that physically inspired wave functions, as the Gutzwiller-projected states~\cite{BeccaGutz2013}, which give a remarkable result with only a few 
parameters, are not always equally accurate in other cases. 

In Fig.~\ref{fig:opt} we show a typical Deep ViT optimization on the $10 \times 10$ lattice at $J_2/J_1=0.5$. First, we optimize the Transformer having translational invariance among 
patches (blue curve). Then, starting from the previously optimized parameters, we restore sequentially the full translational invariance (green curve), rotational symmetry (orange 
curve) and lastly, reflections and spin parity symmetry (red curve). Whenever a new symmetry is restored, the energy reliably decreases~\cite{Nomura_2021}. We stress 
that our optimization process, which combines the SR formulation of Eq.~\eqref{eq:sr2} with a real-valued Deep ViT followed by a complex-valued fully connected output 
layer~\cite{shastry2023}, is highly stable and insensitive to the initial seed, ensuring consistent results across multiple optimization runs.

\section{Discussion.}
We have introduced a formulation of the SR that excels in scenarios where the number of parameters ($P$) significantly outweighs the number of samples ($M$) used for 
stochastic estimations. Exploiting this approach, we attained the state-of-the-art ground state energy for the $J_1$-$J_2$ Heisenberg model at $J_2/J_1=0.5$, on a $10\times10$ 
square lattice, optimizing a Deep ViT with $P=267720$ parameters and using only $M=6000$ samples. It is essential to note that this achievement highlights the remarkable capability of deep 
learning in performing exceptionally well even with a limited sample sizes relative to the overall parameter count. This also challenges the common belief that large 
amount of Monte Carlo samples are required to find the solution in the exponentially-large Hilbert space and for precise SR optimizations~\cite{LiangCNN2022}.

Our results have important ramifications for investigating the physical properties of challenging quantum many-body problems, where the use of the SR is crucial to obtain accurate 
results. The use of large-scale Neural-Network Quantum States can open new opportunities in approximating ground states of quantum spin Hamiltonians, 
where other methods fail. Additionally, making minor modifications to the equations describing parameter updates within the SR framework [see Eq.~\eqref{eq:sr2}] enables us to 
describe the unitary time evolution of quantum many-body systems according to the time-dependent variational principle (TDVP)~\cite{BeccaSorella2017,HeylDyn2023,SchmittDyn2020}. 
Extending our approach to address time-dependent problems stands as a promising avenue for future works. Furthermore, this formulation of the SR can be a key tool for obtaining 
accurate results also in quantum chemistry, especially for systems that suffer from the sign problem~\cite{Nakano_2020}. Typically, in this case, the standard strategy is to 
take $M>10\times P$~\cite{BeccaSorella2017}, which may be unnecessary for deep learning based approaches.

\section{Methods}
\textbf{Linear algebra identity.} 
The key point of the method is the transformation from Eq.~\eqref{eq:sr1} to Eq.~\eqref{eq:sr2}, which uses the matrix identity
\begin{equation}\label{eq:identity}
     (AB + \lambda \mathbb{I}_n)^{-1}A=A(BA + \lambda \mathbb{I}_{m})^{-1} \ ,
\end{equation}
where $A$ and $B$ are $n \times m$ and $m \times n$ matrices, respectively. This identity can be proved starting from
\begin{equation}
    \mathbb{I}_{m}=(BA + \lambda \mathbb{I}_{m})(BA + \lambda \mathbb{I}_{m})^{-1} \ ,
\end{equation}
then, multiplying from the left by $A$, we get
\begin{equation}
    A=A(BA + \lambda \mathbb{I}_{m})(BA + \lambda \mathbb{I}_{m})^{-1} \ ,
\end{equation}
and exploiting the fact that $A \mathbb{I}_{m} = \mathbb{I}_{n} A$, we obtain
\begin{equation} \label{eq:proof}
    A=(AB + \lambda \mathbb{I}_n)A(BA + \lambda \mathbb{I}_{m})^{-1} \ .
\end{equation}
At the end, multiplying from the left by $(AB + \lambda \mathbb{I}_n)^{-1}$, we recover Eq.~\eqref{eq:identity}. 

The identity in Eq.~\eqref{eq:identity} is also used in the \textit{kernel trick}, which is at the basis of kernel methods which have applications in many areas of machine learning~\cite{bishop2006}, including many body quantum systems~\cite{giuliani2023}.\\

\begin{figure}[t]
  \center
  \includegraphics[width=0.9\columnwidth]{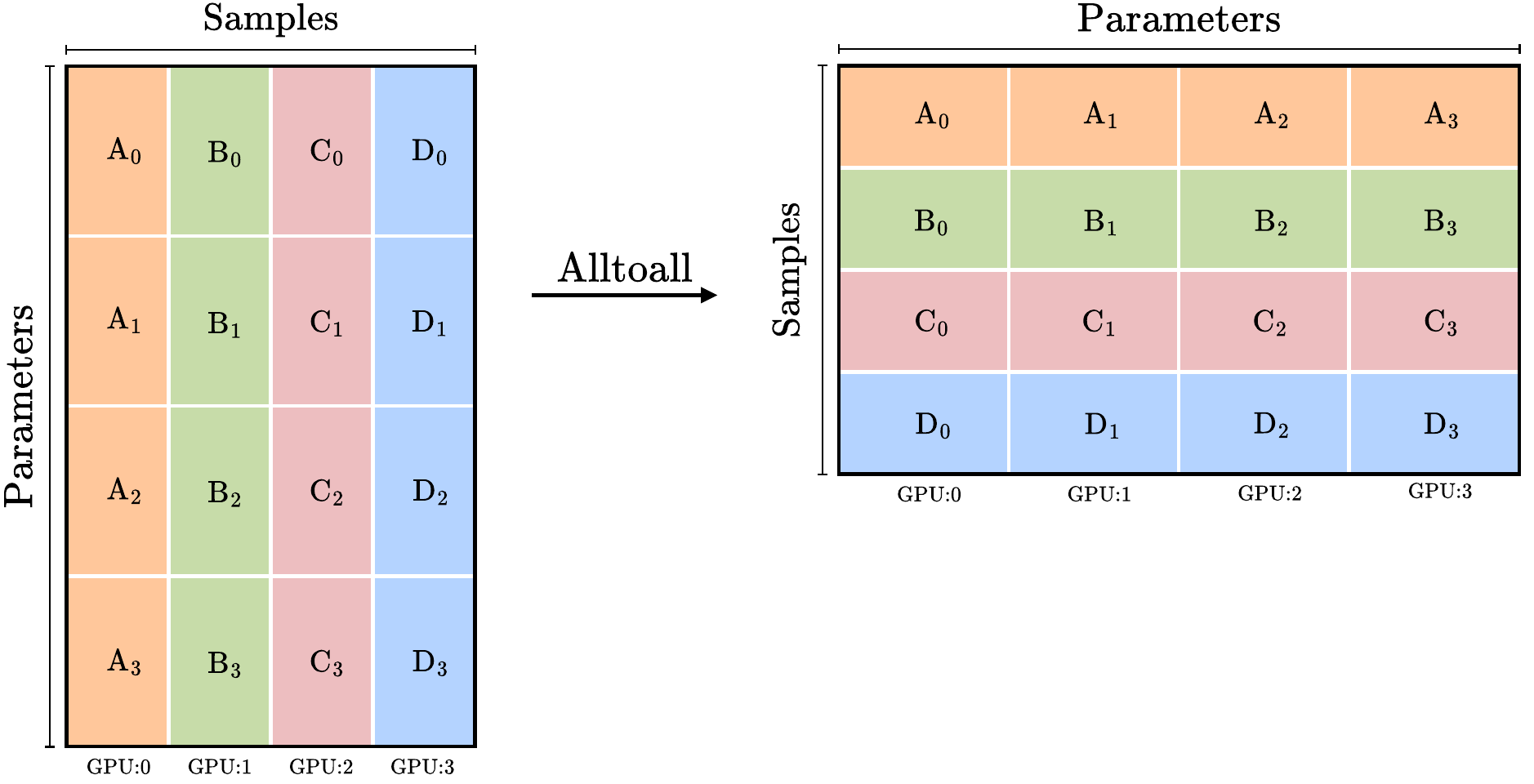}
  \caption{\label{fig:mem} \textbf{Message Passing Interface (MPI) \textit{alltoall} operation.} Graphical representation of MPI \textit{alltoall} operation to transpose the $X$ matrix distributed across multiple GPUs. For example, GPU:0 initially contains sub-matrices $A_0, A_1, A_2, A_3$, while following the transposition, GPU:0 contains sub-matrices $A_0, B_0, C_0, D_0$.}
\end{figure}

\textbf{Distributed SR computation.} The algorithm proposed in Eq.~\eqref{eq:sr2} can be efficiently distributed, both in terms of computational operations and memory, across multiple GPUs. To illustrate this, we 
consider for simplicity the case of a real-valued wave function, where~${X = Y_R \equiv Y}$.
Given a number $M$ of configurations, they can be distributed across $n_G$ GPUs, facilitating parallel simulation of Markov chains. In this way, on the $g$-$th$ GPU, the elements 
$i \in [g M/n_G, (g+1)M/n_G)$ of the vector $\bm{f}$ are obtained, along with the columns $i \in [g M/n_G, (g+1)M/n_G)$ of the matrix $X$, which we indicate using $X_{[:,g]}$. To efficiently apply Eq.~\eqref{eq:sr2}, we employ the 
Message Passing Interface (MPI) \textit{alltoall} collective operation to transpose $X$, yielding the sub-matrix $X_{[g,:]}$ on $g$-th GPU. This sub-matrix comprises the rows elements in $[gP/n_G,(g+1)P/n_G)$ 
of the original matrix $X$ (see Fig.~\ref{fig:mem}). Consequently, we can express:
\begin{equation}
X^T X = \sum_{g=0}^{n_G-1} X_{[g,:]}^T X_{[g,:]} \ .
\end{equation} 

\begin{figure}[t]
  \center
  \includegraphics[width=0.85\columnwidth]{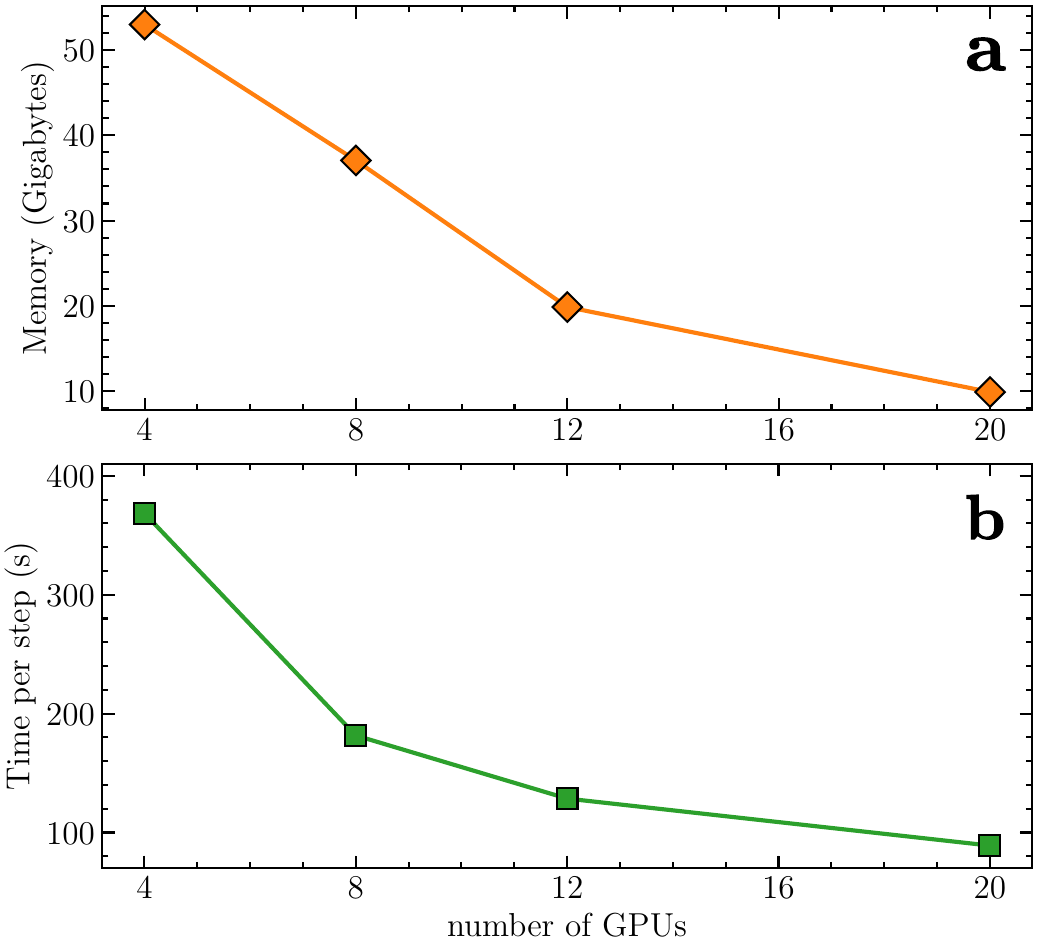}
  \caption{\label{fig:time_mem} \textbf{Memory and time usage.} Memory usage in Gigabytes [see panel \textbf{a}] and computational time per optimization step in seconds [see panel \textbf{b}] as a function of the number of GPUs. The reported values are related to a ViT architeture with $h=12$, $d=72$, $n_l = 8$, fully symmetrized [see Eq.~\eqref{eq:symmViT}] and optimized with $M=6000$ samples.}
\end{figure}

The inner products can be computed in parallel on each GPU, while the outer sum is performed using the MPI primitive \textit{reduce} with the \textit{sum} operation. The 
\textit{master} GPU performs the inversion, computes the vector ${\bm{t} = (X^TX + \lambda \mathbb{I}_{2M})^{-1}\bm{f}}$, and then scatters it across the other GPUs. Finally, after transposing again the matrix $X$ with the MPI \textit{alltoall} operation, the parameter update can be computed as follows:
\begin{equation}
    \bm{\delta \theta} = \tau \sum_{g=0}^{n_G-1} X_{[:,g]}  \bm{t}_{g} \ .
\end{equation}
This procedure significantly reduces the memory requirement per GPU to $O(MP/n_G)$, enabling the optimization of an arbitrary number of parameters using the SR approach. In Fig.~\ref{fig:time_mem} we report the memory usage and the computational time per optimization step.\\
\begin{figure}[b]
  \center
  \includegraphics[width=\columnwidth]{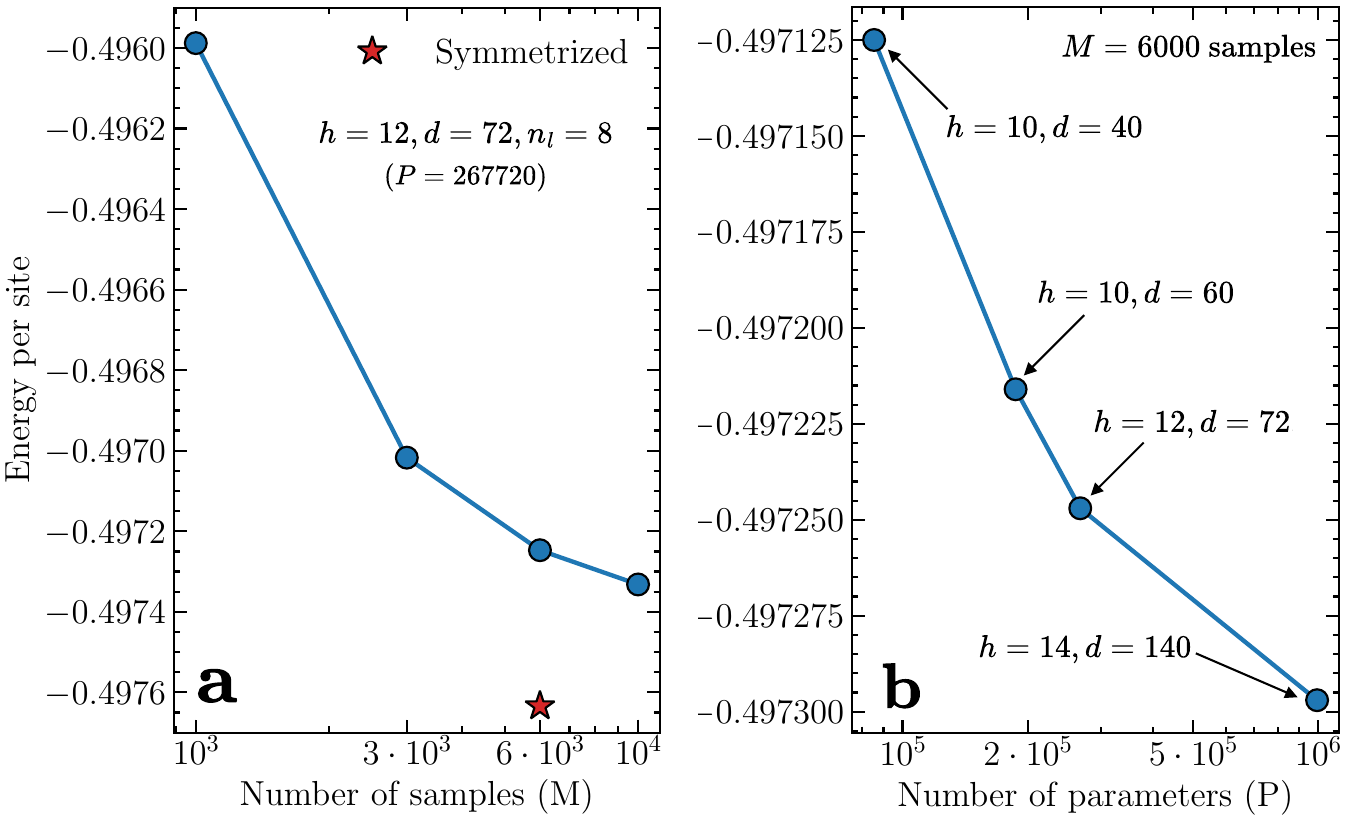}
  \caption{\label{fig:scaling_en} \textbf{Variational energy scalings.} Panel \textbf{a}: Energy per site as a function of the number of samples $M$ for a ViT with $n_l=8$ layers, embedding dimension $d=72$ and $h=12$ heads per layer. Panel \textbf{b}: Energy per site as a function of the number of parameters $P$, increased by adding heads $h$ and taking larger embedding dimensions $d$, for a fixed number of layers $n_l=8$. For both panels, the energy values (blue circles) are obtained without restoring the symmetries; for comparison we also show the energy corresponding to the fully symmetrized state in Eq.~\eqref{eq:symmViT} (red star) which is the one reported in Table~\ref{table:accuracy}.}
\end{figure}

\textbf{Systematic energy improvement.} Here, we discuss the impact of the number of samples $M$ and parameters $P$ on the variational energy of the ViT wave function, showing the results in Fig.~\ref{fig:scaling_en}.
In the right panel, we show the variational energy as a function of the number of parameters for a fixed number of layers $n_l=8$, performing the optimizations with $M=6000$ samples. The number of parameters is increased by enlarging the width of each layer. In particular, we take the following architectures: $(h=10, d=40)$, $(h=10, d=60)$, $(h=12, d=72)$, and $(h=14, d=140)$ with $P=85400$, $187100$, $267720$, and $994700$ parameters, respectively. A similar analysis for a fixed width but varying the number of layers is discussed in Ref.~\cite{shastry2023}. Instead, in the left panel, we fix an architecture $(h=12, d=72)$ with $n_l=8$ and increase the number of samples $M$ up to $10^4$. Both analyses are performed without restoring the symmetries by quantum number projection; for comparison, we report in the left panel the energy obtained after restoring the symmetries [see Eq.~\eqref{eq:symmViT}]. The latter one coincides with the ViT wave function used to obtain the energy reported in Table~\ref{table:accuracy}. We point out that the energy curves depicted in Fig.~\ref{fig:scaling_en} are obtained from unbiased simulations, without the utilization of Marshall sign prior. The final value of the energy, as well as the convergence to it, are qualitatively similar regardless of its inclusion. \\

\textbf{Details on the Transformer architecture.} In this section we provide a pseudocode (see Algorithm~\ref{algorithm}) describing the steps for the implementation of the Vision Transformer architecture employed in this work and described in \textit{Results}. In particular we emphasize that skip connections and Layer Normalization are implemented as described in Ref.~\cite{xiong2020layer}. 

\begin{algorithm}[H]
  \begin{algorithmic}[1]
       \State Input configuration $\sigma \in \{-1,1\}^{L \times L}$
      \State Patch and Embed: $\mathcal{X} \leftarrow (\boldsymbol{x}_1, \dots \boldsymbol{x}_{L^2/b^2}) \in \mathbb{R}^{\nicefrac{L^2}{b^2} \times d}$
      \For{$i = 1, n_l$} \label{forloop}
        \State $\mathcal{X} \leftarrow \mathcal{X} + \text{MHFA}(\text{LayerNorm($\mathcal{X}$)})$
        \State $\mathcal{X} \leftarrow \mathcal{X} + \text{FFN}(\text{LayerNorm($\mathcal{X}$)})$
      \EndFor
      \State $(\boldsymbol{y}_1, \dots \boldsymbol{y}_{L^2/b^2}) \leftarrow \text{LayerNorm}(\mathcal{X})$
      \State $\boldsymbol{z} \leftarrow \sum_{i=1}^d \boldsymbol{y}_{i}$
      \State $\text{Log}[\Psi_{\theta}(\sigma)] \leftarrow \sum_{\alpha=1}^d g(b_{\alpha} + \boldsymbol{w}_{\alpha}\cdot\boldsymbol{z})$
  \end{algorithmic}
  \caption{Vision Transformer Wave Function} \label{algorithm}
\end{algorithm}

\textbf{Monte Carlo sampling.}
The expectation value of an operator $\hat{A}$ on a given variational state $\ket{\Psi_{\theta}}$ can be computed as
\begin{equation}\label{eq:exp_values}
    \braket{\hat{A}} = \frac{\braket{\Psi_{\theta}|\hat{A}|\Psi_{\theta}}}{\braket{\Psi_{\theta}|\Psi_{\theta}}} = \sum_{\{\sigma\}} P_{\theta}(\sigma) A_L(\sigma) \ ,
\end{equation}
where $P_{\theta}(\sigma) \propto |\Psi_{\theta}(\sigma)|^2$ and $A_L(\sigma) = \braket{\sigma|\hat{A}|\Psi_{\theta}}/\braket{\sigma|\Psi_{\theta}}$ is the so-called local estimator of $\hat{A}$
\begin{equation}
    A_L(\sigma) = \sum_{\{\sigma'\}} \braket{\sigma|\hat{A}|\sigma'}\frac{\Psi_{\theta}(\sigma')}{\Psi_{\theta}(\sigma)} \ .
\end{equation}
For any local operator (e.g., the Hamiltonian) the matrix $\braket{\sigma|\hat{A}|\sigma'}$ is sparse, then the calculation of $A_L(\sigma)$ is at most polynomial in the number of spins. Furthermore, if it is possible to efficiently generate  a sample of configurations $\{\sigma_1, \sigma_2, \dots, \sigma_M\}$ from the distribution $P_{\theta}(\sigma)$
(e.g., by performing a Markov chain Monte Carlo), then Eq.~\eqref{eq:exp_values} can be used to obtain a stochastic estimation of the expectation value \\
\begin{equation}
     \bar{A} =\frac{1}{M}\sum_{i = 1}^{M} A_L(\sigma_i) \ ,
\end{equation}
where $\bar{A} \approx \braket{\hat{A}}$ and the accuracy of the estimation is controlled by a statistical error which scales as $O(1/\sqrt{M})$.\\

\textit{Note added to the proof.} During the revision process, we became aware of an updated version of Ref.~\cite{HeylCNN2023} where a variational energy per site of -0.4976921(4) has been obtained.\\

\textbf{Data availability.} The data that support the findings of this study are available from the corresponding
author upon reasonable request. \\

\textbf{Code availability.} The variational quantum Monte Carlo and the Deep ViT architecture were implemented in JAX~\cite{jax2018github}. The parallel implementation of the Stochastic Reconfiguration was 
implemented using mpi4jax~\cite{mpi4jax}, and it is available on NetKet~\cite{netket3}, under the name of
\href{https://netket.readthedocs.io/en/latest/api/_generated/experimental/driver/netket.experimental.driver.VMC_SRt.html#netket.experimental.driver.VMC_SRt}{\texttt{VMC\_SRt}}. The ViT architecture will be made available from the authors upon reasonable request. \\

\textbf{Author contributions} R.R., L.L.V. and L.B. devised the algorithm with input from F.B. and S.G., and performed the numerical simulations. R.R., L.L.V., L.B., F.B. and S.G. wrote the manuscript. \\

\textbf{Competing interests.} The authors declare no competing interests. \\

\begin{acknowledgments}
We thank R.\ Favata for critically reading the manuscript and M.\ Imada for stimulating us with challenging questions. We also acknowledge G.\ Carleo, F.\ Vicentini, Y.\ Nomura, A.\ Szabó, J.\ Carrasquilla and A.\ Chen for useful discussions.
The variational quantum Monte Carlo and the Deep ViT architecture were implemented in JAX~\cite{jax2018github}. The parallel implementation of the Stochastic Reconfiguration was 
implemented using mpi4jax~\cite{mpi4jax}, and it is available on NetKet~\cite{netket3}, under the name of
\href{https://netket.readthedocs.io/en/latest/api/_generated/experimental/driver/netket.experimental.driver.VMC_SRt.html#netket.experimental.driver.VMC_SRt}{\texttt{VMC\_SRt}}. R.R.~and L.L.V.~acknowledge the CINECA award under the ISCRA initiative, for the availability of high-performance computing resources and support. S.G.~acknowledges co-funding from Next Generation EU, in the context of the National Recovery and Resilience Plan, Investment PE1 – Project FAIR “Future Artificial Intelligence Research”. This resource was 
co-financed by the Next Generation EU [DM 1555 del 11.10.22]. \\
\end{acknowledgments}

\bibliography{refs}

\begin{thebibliography}{65}%
\makeatletter
\providecommand \@ifxundefined [1]{%
 \@ifx{#1\undefined}
}%
\providecommand \@ifnum [1]{%
 \ifnum #1\expandafter \@firstoftwo
 \else \expandafter \@secondoftwo
 \fi
}%
\providecommand \@ifx [1]{%
 \ifx #1\expandafter \@firstoftwo
 \else \expandafter \@secondoftwo
 \fi
}%
\providecommand \natexlab [1]{#1}%
\providecommand \enquote  [1]{``#1''}%
\providecommand \bibnamefont  [1]{#1}%
\providecommand \bibfnamefont [1]{#1}%
\providecommand \citenamefont [1]{#1}%
\providecommand \href@noop [0]{\@secondoftwo}%
\providecommand \href [0]{\begingroup \@sanitize@url \@href}%
\providecommand \@href[1]{\@@startlink{#1}\@@href}%
\providecommand \@@href[1]{\endgroup#1\@@endlink}%
\providecommand \@sanitize@url [0]{\catcode `\\12\catcode `\$12\catcode
  `\&12\catcode `\#12\catcode `\^12\catcode `\_12\catcode `\%12\relax}%
\providecommand \@@startlink[1]{}%
\providecommand \@@endlink[0]{}%
\providecommand \url  [0]{\begingroup\@sanitize@url \@url }%
\providecommand \@url [1]{\endgroup\@href {#1}{\urlprefix }}%
\providecommand \urlprefix  [0]{URL }%
\providecommand \Eprint [0]{\href }%
\providecommand \doibase [0]{http://dx.doi.org/}%
\providecommand \selectlanguage [0]{\@gobble}%
\providecommand \bibinfo  [0]{\@secondoftwo}%
\providecommand \bibfield  [0]{\@secondoftwo}%
\providecommand \translation [1]{[#1]}%
\providecommand \BibitemOpen [0]{}%
\providecommand \bibitemStop [0]{}%
\providecommand \bibitemNoStop [0]{.\EOS\space}%
\providecommand \EOS [0]{\spacefactor3000\relax}%
\providecommand \BibitemShut  [1]{\csname bibitem#1\endcsname}%
\let\auto@bib@innerbib\@empty
\bibitem [{\citenamefont {Krizhevsky}\ \emph {et~al.}(2012)\citenamefont
  {Krizhevsky}, \citenamefont {Sutskever},\ and\ \citenamefont
  {Hinton}}]{imagenet2012}%
  \BibitemOpen
  \bibfield  {author} {\bibinfo {author} {\bibfnamefont {A.}~\bibnamefont
  {Krizhevsky}}, \bibinfo {author} {\bibfnamefont {I.}~\bibnamefont
  {Sutskever}}, \ and\ \bibinfo {author} {\bibfnamefont {G.~E.}\ \bibnamefont
  {Hinton}},\ }in\ \href
  {https://proceedings.neurips.cc/paper_files/paper/2012/file/c399862d3b9d6b76c8436e924a68c45b-Paper.pdf}
  {\emph {\bibinfo {booktitle} {Advances in Neural Information Processing
  Systems}}},\ Vol.~\bibinfo {volume} {25},\ \bibinfo {editor} {edited by\
  \bibinfo {editor} {\bibfnamefont {F.}~\bibnamefont {Pereira}}, \bibinfo
  {editor} {\bibfnamefont {C.}~\bibnamefont {Burges}}, \bibinfo {editor}
  {\bibfnamefont {L.}~\bibnamefont {Bottou}}, \ and\ \bibinfo {editor}
  {\bibfnamefont {K.}~\bibnamefont {Weinberger}}}\ (\bibinfo  {publisher}
  {Curran Associates, Inc.},\ \bibinfo {year} {2012})\BibitemShut {NoStop}%
\bibitem [{\citenamefont {He}\ \emph {et~al.}(2016)\citenamefont {He},
  \citenamefont {Zhang}, \citenamefont {Ren},\ and\ \citenamefont
  {Sun}}]{residuals2016}%
  \BibitemOpen
  \bibfield  {author} {\bibinfo {author} {\bibfnamefont {K.}~\bibnamefont
  {He}}, \bibinfo {author} {\bibfnamefont {X.}~\bibnamefont {Zhang}}, \bibinfo
  {author} {\bibfnamefont {S.}~\bibnamefont {Ren}}, \ and\ \bibinfo {author}
  {\bibfnamefont {J.}~\bibnamefont {Sun}},\ }in\ \href {\doibase
  10.1109/CVPR.2016.90} {\emph {\bibinfo {booktitle} {2016 IEEE Conference on
  Computer Vision and Pattern Recognition (CVPR)}}}\ (\bibinfo {year} {2016})\
  pp.\ \bibinfo {pages} {770--778}\BibitemShut {NoStop}%
\bibitem [{\citenamefont {Vaswani}\ \emph {et~al.}(2017)\citenamefont
  {Vaswani}, \citenamefont {Shazeer}, \citenamefont {Parmar}, \citenamefont
  {Uszkoreit}, \citenamefont {Jones}, \citenamefont {Gomez}, \citenamefont
  {Kaiser},\ and\ \citenamefont {Polosukhin}}]{vaswani2017attention}%
  \BibitemOpen
  \bibfield  {author} {\bibinfo {author} {\bibfnamefont {A.}~\bibnamefont
  {Vaswani}}, \bibinfo {author} {\bibfnamefont {N.}~\bibnamefont {Shazeer}},
  \bibinfo {author} {\bibfnamefont {N.}~\bibnamefont {Parmar}}, \bibinfo
  {author} {\bibfnamefont {J.}~\bibnamefont {Uszkoreit}}, \bibinfo {author}
  {\bibfnamefont {L.}~\bibnamefont {Jones}}, \bibinfo {author} {\bibfnamefont
  {A.~N.}\ \bibnamefont {Gomez}}, \bibinfo {author} {\bibfnamefont
  {{\L}.}~\bibnamefont {Kaiser}}, \ and\ \bibinfo {author} {\bibfnamefont
  {I.}~\bibnamefont {Polosukhin}},\ }\href@noop {} {\bibfield  {journal}
  {\bibinfo  {journal} {Advances in neural information processing systems}\
  }\textbf {\bibinfo {volume} {30}} (\bibinfo {year} {2017})}\BibitemShut
  {NoStop}%
\bibitem [{\citenamefont {Devlin}\ \emph {et~al.}(2019)\citenamefont {Devlin},
  \citenamefont {Chang}, \citenamefont {Lee},\ and\ \citenamefont
  {Toutanova}}]{bert2019}%
  \BibitemOpen
  \bibfield  {author} {\bibinfo {author} {\bibfnamefont {J.}~\bibnamefont
  {Devlin}}, \bibinfo {author} {\bibfnamefont {M.-W.}\ \bibnamefont {Chang}},
  \bibinfo {author} {\bibfnamefont {K.}~\bibnamefont {Lee}}, \ and\ \bibinfo
  {author} {\bibfnamefont {K.}~\bibnamefont {Toutanova}},\ }\href@noop {}
  {\enquote {\bibinfo {title} {Bert: Pre-training of deep bidirectional
  transformers for language understanding},}\ } (\bibinfo {year} {2019}),\
  \Eprint {http://arxiv.org/abs/1810.04805} {arXiv:1810.04805 [cs.CL]}
  \BibitemShut {NoStop}%
\bibitem [{\citenamefont {Brown}\ \emph {et~al.}(2020)\citenamefont {Brown},
  \citenamefont {Mann}, \citenamefont {Ryder} \emph {et~al.}}]{nlp2020}%
  \BibitemOpen
  \bibfield  {author} {\bibinfo {author} {\bibfnamefont {T.}~\bibnamefont
  {Brown}}, \bibinfo {author} {\bibfnamefont {B.}~\bibnamefont {Mann}},
  \bibinfo {author} {\bibfnamefont {N.}~\bibnamefont {Ryder}},  \emph
  {et~al.},\ }in\ \href
  {https://proceedings.neurips.cc/paper_files/paper/2020/file/1457c0d6bfcb4967418bfb8ac142f64a-Paper.pdf}
  {\emph {\bibinfo {booktitle} {Advances in Neural Information Processing
  Systems}}},\ Vol.~\bibinfo {volume} {33}\ (\bibinfo  {publisher} {Curran
  Associates, Inc.},\ \bibinfo {year} {2020})\ pp.\ \bibinfo {pages}
  {1877--1901}\BibitemShut {NoStop}%
\bibitem [{\citenamefont {Kingma}\ and\ \citenamefont {Ba}(2017)}]{adam2017}%
  \BibitemOpen
  \bibfield  {author} {\bibinfo {author} {\bibfnamefont {D.~P.}\ \bibnamefont
  {Kingma}}\ and\ \bibinfo {author} {\bibfnamefont {J.}~\bibnamefont {Ba}},\
  }\href@noop {} {\enquote {\bibinfo {title} {Adam: A method for stochastic
  optimization},}\ } (\bibinfo {year} {2017}),\ \Eprint
  {http://arxiv.org/abs/1412.6980} {arXiv:1412.6980 [cs.LG]} \BibitemShut
  {NoStop}%
\bibitem [{\citenamefont {Loshchilov}\ and\ \citenamefont
  {Hutter}(2019)}]{adamw2019}%
  \BibitemOpen
  \bibfield  {author} {\bibinfo {author} {\bibfnamefont {I.}~\bibnamefont
  {Loshchilov}}\ and\ \bibinfo {author} {\bibfnamefont {F.}~\bibnamefont
  {Hutter}},\ }\href@noop {} {\enquote {\bibinfo {title} {Decoupled weight
  decay regularization},}\ } (\bibinfo {year} {2019}),\ \Eprint
  {http://arxiv.org/abs/1711.05101} {arXiv:1711.05101 [cs.LG]} \BibitemShut
  {NoStop}%
\bibitem [{\citenamefont {Amari}\ and\ \citenamefont
  {Douglas}(1998)}]{Amari1998}%
  \BibitemOpen
  \bibfield  {author} {\bibinfo {author} {\bibfnamefont {S.}~\bibnamefont
  {Amari}}\ and\ \bibinfo {author} {\bibfnamefont {S.}~\bibnamefont
  {Douglas}},\ }in\ \href {\doibase 10.1109/ICASSP.1998.675489} {\emph
  {\bibinfo {booktitle} {Proceedings of the 1998 IEEE International Conference
  on Acoustics, Speech and Signal Processing, ICASSP '98 (Cat.
  No.98CH36181)}}},\ Vol.~\bibinfo {volume} {2}\ (\bibinfo {year} {1998})\ pp.\
  \bibinfo {pages} {1213--1216 vol.2}\BibitemShut {NoStop}%
\bibitem [{\citenamefont {Amari}\ \emph {et~al.}(2019)\citenamefont {Amari},
  \citenamefont {Karakida},\ and\ \citenamefont {Oizumi}}]{Amari2018}%
  \BibitemOpen
  \bibfield  {author} {\bibinfo {author} {\bibfnamefont {S.}~\bibnamefont
  {Amari}}, \bibinfo {author} {\bibfnamefont {R.}~\bibnamefont {Karakida}}, \
  and\ \bibinfo {author} {\bibfnamefont {M.}~\bibnamefont {Oizumi}},\ }in\
  \href {https://proceedings.mlr.press/v89/amari19a.html} {\emph {\bibinfo
  {booktitle} {Proceedings of the Twenty-Second International Conference on
  Artificial Intelligence and Statistics}}},\ \bibinfo {series} {Proceedings of
  Machine Learning Research}, Vol.~\bibinfo {volume} {89},\ \bibinfo {editor}
  {edited by\ \bibinfo {editor} {\bibfnamefont {K.}~\bibnamefont {Chaudhuri}}\
  and\ \bibinfo {editor} {\bibfnamefont {M.}~\bibnamefont {Sugiyama}}}\
  (\bibinfo  {publisher} {PMLR},\ \bibinfo {year} {2019})\ pp.\ \bibinfo
  {pages} {694--702}\BibitemShut {NoStop}%
\bibitem [{\citenamefont {Sorella}(1998)}]{Sorella1998}%
  \BibitemOpen
  \bibfield  {author} {\bibinfo {author} {\bibfnamefont {S.}~\bibnamefont
  {Sorella}},\ }\href {\doibase 10.1103/PhysRevLett.80.4558} {\bibfield
  {journal} {\bibinfo  {journal} {Phys. Rev. Lett.}\ }\textbf {\bibinfo
  {volume} {80}},\ \bibinfo {pages} {4558} (\bibinfo {year}
  {1998})}\BibitemShut {NoStop}%
\bibitem [{\citenamefont {Sorella}(2005)}]{Sorella2005}%
  \BibitemOpen
  \bibfield  {author} {\bibinfo {author} {\bibfnamefont {S.}~\bibnamefont
  {Sorella}},\ }\href {\doibase 10.1103/PhysRevB.71.241103} {\bibfield
  {journal} {\bibinfo  {journal} {Phys. Rev. B}\ }\textbf {\bibinfo {volume}
  {71}},\ \bibinfo {pages} {241103} (\bibinfo {year} {2005})}\BibitemShut
  {NoStop}%
\bibitem [{\citenamefont {Park}\ and\ \citenamefont
  {Kastoryano}(2020)}]{Park2020}%
  \BibitemOpen
  \bibfield  {author} {\bibinfo {author} {\bibfnamefont {C.~Y.}\ \bibnamefont
  {Park}}\ and\ \bibinfo {author} {\bibfnamefont {M.~J.}\ \bibnamefont
  {Kastoryano}},\ }\href {\doibase 10.1103/PhysRevResearch.2.023232} {\bibfield
   {journal} {\bibinfo  {journal} {Phys. Rev. Res.}\ }\textbf {\bibinfo
  {volume} {2}},\ \bibinfo {pages} {023232} (\bibinfo {year}
  {2020})}\BibitemShut {NoStop}%
\bibitem [{\citenamefont {Capello}\ \emph {et~al.}(2005)\citenamefont
  {Capello}, \citenamefont {Becca}, \citenamefont {Fabrizio}, \citenamefont
  {Sorella},\ and\ \citenamefont {Tosatti}}]{Capello2005}%
  \BibitemOpen
  \bibfield  {author} {\bibinfo {author} {\bibfnamefont {M.}~\bibnamefont
  {Capello}}, \bibinfo {author} {\bibfnamefont {F.}~\bibnamefont {Becca}},
  \bibinfo {author} {\bibfnamefont {M.}~\bibnamefont {Fabrizio}}, \bibinfo
  {author} {\bibfnamefont {S.}~\bibnamefont {Sorella}}, \ and\ \bibinfo
  {author} {\bibfnamefont {E.}~\bibnamefont {Tosatti}},\ }\href {\doibase
  10.1103/PhysRevLett.94.026406} {\bibfield  {journal} {\bibinfo  {journal}
  {Phys. Rev. Lett.}\ }\textbf {\bibinfo {volume} {94}},\ \bibinfo {pages}
  {026406} (\bibinfo {year} {2005})}\BibitemShut {NoStop}%
\bibitem [{\citenamefont {Hu}\ \emph {et~al.}(2013)\citenamefont {Hu},
  \citenamefont {Becca}, \citenamefont {Parola},\ and\ \citenamefont
  {Sorella}}]{BeccaGutz2013}%
  \BibitemOpen
  \bibfield  {author} {\bibinfo {author} {\bibfnamefont {W.-J.}\ \bibnamefont
  {Hu}}, \bibinfo {author} {\bibfnamefont {F.}~\bibnamefont {Becca}}, \bibinfo
  {author} {\bibfnamefont {A.}~\bibnamefont {Parola}}, \ and\ \bibinfo {author}
  {\bibfnamefont {S.}~\bibnamefont {Sorella}},\ }\href {\doibase
  10.1103/PhysRevB.88.060402} {\bibfield  {journal} {\bibinfo  {journal} {Phys.
  Rev. B}\ }\textbf {\bibinfo {volume} {88}},\ \bibinfo {pages} {060402}
  (\bibinfo {year} {2013})}\BibitemShut {NoStop}%
\bibitem [{\citenamefont {Carleo}\ and\ \citenamefont
  {Troyer}(2017)}]{Carleo2017}%
  \BibitemOpen
  \bibfield  {author} {\bibinfo {author} {\bibfnamefont {G.}~\bibnamefont
  {Carleo}}\ and\ \bibinfo {author} {\bibfnamefont {M.}~\bibnamefont
  {Troyer}},\ }\href {\doibase 10.1126/science.aag2302} {\bibfield  {journal}
  {\bibinfo  {journal} {Science}\ }\textbf {\bibinfo {volume} {355}},\ \bibinfo
  {pages} {602} (\bibinfo {year} {2017})}\BibitemShut {NoStop}%
\bibitem [{\citenamefont {Ferrari}\ \emph {et~al.}(2019)\citenamefont
  {Ferrari}, \citenamefont {Becca},\ and\ \citenamefont
  {Carrasquilla}}]{FerrariGutz2019}%
  \BibitemOpen
  \bibfield  {author} {\bibinfo {author} {\bibfnamefont {F.}~\bibnamefont
  {Ferrari}}, \bibinfo {author} {\bibfnamefont {F.}~\bibnamefont {Becca}}, \
  and\ \bibinfo {author} {\bibfnamefont {J.}~\bibnamefont {Carrasquilla}},\
  }\href {\doibase 10.1103/PhysRevB.100.125131} {\bibfield  {journal} {\bibinfo
   {journal} {Phys. Rev. B}\ }\textbf {\bibinfo {volume} {100}},\ \bibinfo
  {pages} {125131} (\bibinfo {year} {2019})}\BibitemShut {NoStop}%
\bibitem [{\citenamefont {Nomura}\ \emph {et~al.}(2017)\citenamefont {Nomura},
  \citenamefont {Darmawan}, \citenamefont {Yamaji},\ and\ \citenamefont
  {Imada}}]{Nomura2017}%
  \BibitemOpen
  \bibfield  {author} {\bibinfo {author} {\bibfnamefont {Y.}~\bibnamefont
  {Nomura}}, \bibinfo {author} {\bibfnamefont {A.~S.}\ \bibnamefont
  {Darmawan}}, \bibinfo {author} {\bibfnamefont {Y.}~\bibnamefont {Yamaji}}, \
  and\ \bibinfo {author} {\bibfnamefont {M.}~\bibnamefont {Imada}},\ }\href
  {\doibase 10.1103/PhysRevB.96.205152} {\bibfield  {journal} {\bibinfo
  {journal} {Phys. Rev. B}\ }\textbf {\bibinfo {volume} {96}},\ \bibinfo
  {pages} {205152} (\bibinfo {year} {2017})}\BibitemShut {NoStop}%
\bibitem [{\citenamefont {Viteritti}\ \emph {et~al.}(2022)\citenamefont
  {Viteritti}, \citenamefont {Ferrari},\ and\ \citenamefont
  {Becca}}]{Viteritti2022}%
  \BibitemOpen
  \bibfield  {author} {\bibinfo {author} {\bibfnamefont {L.}~\bibnamefont
  {Viteritti}}, \bibinfo {author} {\bibfnamefont {F.}~\bibnamefont {Ferrari}},
  \ and\ \bibinfo {author} {\bibfnamefont {F.}~\bibnamefont {Becca}},\ }\href
  {\doibase 10.21468/SciPostPhys.12.5.166} {\bibfield  {journal} {\bibinfo
  {journal} {SciPost Phys.}\ }\textbf {\bibinfo {volume} {12}},\ \bibinfo
  {pages} {166} (\bibinfo {year} {2022})}\BibitemShut {NoStop}%
\bibitem [{\citenamefont {Park}\ and\ \citenamefont
  {Kastoryano}(2022)}]{Park2022}%
  \BibitemOpen
  \bibfield  {author} {\bibinfo {author} {\bibfnamefont {C.-Y.}\ \bibnamefont
  {Park}}\ and\ \bibinfo {author} {\bibfnamefont {M.~J.}\ \bibnamefont
  {Kastoryano}},\ }\href {\doibase 10.1103/PhysRevB.106.134437} {\bibfield
  {journal} {\bibinfo  {journal} {Phys. Rev. B}\ }\textbf {\bibinfo {volume}
  {106}},\ \bibinfo {pages} {134437} (\bibinfo {year} {2022})}\BibitemShut
  {NoStop}%
\bibitem [{\citenamefont {Nomura}(2023)}]{Nomura2023}%
  \BibitemOpen
  \bibfield  {author} {\bibinfo {author} {\bibfnamefont {Y.}~\bibnamefont
  {Nomura}},\ }\href@noop {} {\enquote {\bibinfo {title} {Boltzmann machines
  and quantum many-body problems},}\ } (\bibinfo {year} {2023}),\ \Eprint
  {http://arxiv.org/abs/2306.16877} {arXiv:2306.16877 [cond-mat.str-el]}
  \BibitemShut {NoStop}%
\bibitem [{\citenamefont {Choo}\ \emph {et~al.}(2019)\citenamefont {Choo},
  \citenamefont {Neupert},\ and\ \citenamefont {Carleo}}]{CNNCarleo2019}%
  \BibitemOpen
  \bibfield  {author} {\bibinfo {author} {\bibfnamefont {K.}~\bibnamefont
  {Choo}}, \bibinfo {author} {\bibfnamefont {T.}~\bibnamefont {Neupert}}, \
  and\ \bibinfo {author} {\bibfnamefont {G.}~\bibnamefont {Carleo}},\ }\href
  {\doibase 10.1103/PhysRevB.100.125124} {\bibfield  {journal} {\bibinfo
  {journal} {Phys. Rev. B}\ }\textbf {\bibinfo {volume} {100}},\ \bibinfo
  {pages} {125124} (\bibinfo {year} {2019})}\BibitemShut {NoStop}%
\bibitem [{\citenamefont {Liang}\ \emph {et~al.}(2018)\citenamefont {Liang},
  \citenamefont {Liu}, \citenamefont {Lin}, \citenamefont {Guo}, \citenamefont
  {Zhang},\ and\ \citenamefont {He}}]{CNNXiao2018}%
  \BibitemOpen
  \bibfield  {author} {\bibinfo {author} {\bibfnamefont {X.}~\bibnamefont
  {Liang}}, \bibinfo {author} {\bibfnamefont {W.-Y.}\ \bibnamefont {Liu}},
  \bibinfo {author} {\bibfnamefont {P.-Z.}\ \bibnamefont {Lin}}, \bibinfo
  {author} {\bibfnamefont {G.-C.}\ \bibnamefont {Guo}}, \bibinfo {author}
  {\bibfnamefont {Y.-S.}\ \bibnamefont {Zhang}}, \ and\ \bibinfo {author}
  {\bibfnamefont {L.}~\bibnamefont {He}},\ }\href {\doibase
  10.1103/PhysRevB.98.104426} {\bibfield  {journal} {\bibinfo  {journal} {Phys.
  Rev. B}\ }\textbf {\bibinfo {volume} {98}},\ \bibinfo {pages} {104426}
  (\bibinfo {year} {2018})}\BibitemShut {NoStop}%
\bibitem [{\citenamefont {Szab\'o}\ and\ \citenamefont
  {Castelnovo}(2020)}]{CNNSzabo2020}%
  \BibitemOpen
  \bibfield  {author} {\bibinfo {author} {\bibfnamefont {A.}~\bibnamefont
  {Szab\'o}}\ and\ \bibinfo {author} {\bibfnamefont {C.}~\bibnamefont
  {Castelnovo}},\ }\href {\doibase 10.1103/PhysRevResearch.2.033075} {\bibfield
   {journal} {\bibinfo  {journal} {Phys. Rev. Res.}\ }\textbf {\bibinfo
  {volume} {2}},\ \bibinfo {pages} {033075} (\bibinfo {year}
  {2020})}\BibitemShut {NoStop}%
\bibitem [{\citenamefont {Hibat-Allah}\ \emph {et~al.}(2020)\citenamefont
  {Hibat-Allah}, \citenamefont {Ganahl}, \citenamefont {Hayward}, \citenamefont
  {Melko},\ and\ \citenamefont {Carrasquilla}}]{RNNCarrasquilla2020}%
  \BibitemOpen
  \bibfield  {author} {\bibinfo {author} {\bibfnamefont {M.}~\bibnamefont
  {Hibat-Allah}}, \bibinfo {author} {\bibfnamefont {M.}~\bibnamefont {Ganahl}},
  \bibinfo {author} {\bibfnamefont {L.~E.}\ \bibnamefont {Hayward}}, \bibinfo
  {author} {\bibfnamefont {R.~G.}\ \bibnamefont {Melko}}, \ and\ \bibinfo
  {author} {\bibfnamefont {J.}~\bibnamefont {Carrasquilla}},\ }\href {\doibase
  10.1103/PhysRevResearch.2.023358} {\bibfield  {journal} {\bibinfo  {journal}
  {Phys. Rev. Res.}\ }\textbf {\bibinfo {volume} {2}},\ \bibinfo {pages}
  {023358} (\bibinfo {year} {2020})}\BibitemShut {NoStop}%
\bibitem [{\citenamefont {Roth}(2020)}]{RNNRoth2020}%
  \BibitemOpen
  \bibfield  {author} {\bibinfo {author} {\bibfnamefont {C.}~\bibnamefont
  {Roth}},\ }\href@noop {} {\enquote {\bibinfo {title} {Iterative retraining of
  quantum spin models using recurrent neural networks},}\ } (\bibinfo {year}
  {2020}),\ \Eprint {http://arxiv.org/abs/2003.06228} {arXiv:2003.06228
  [physics.comp-ph]} \BibitemShut {NoStop}%
\bibitem [{\citenamefont {Hibat-Allah}\ \emph {et~al.}(2022)\citenamefont
  {Hibat-Allah}, \citenamefont {Melko},\ and\ \citenamefont
  {Carrasquilla}}]{RNNHibatallah2022}%
  \BibitemOpen
  \bibfield  {author} {\bibinfo {author} {\bibfnamefont {M.}~\bibnamefont
  {Hibat-Allah}}, \bibinfo {author} {\bibfnamefont {R.~G.}\ \bibnamefont
  {Melko}}, \ and\ \bibinfo {author} {\bibfnamefont {J.}~\bibnamefont
  {Carrasquilla}},\ }\href@noop {} {\enquote {\bibinfo {title} {Supplementing
  recurrent neural network wave functions with symmetry and annealing to
  improve accuracy},}\ } (\bibinfo {year} {2022}),\ \Eprint
  {http://arxiv.org/abs/2207.14314} {arXiv:2207.14314 [cond-mat.dis-nn]}
  \BibitemShut {NoStop}%
\bibitem [{\citenamefont {Hibat-Allah}\ \emph {et~al.}(2023)\citenamefont
  {Hibat-Allah}, \citenamefont {Melko},\ and\ \citenamefont
  {Carrasquilla}}]{RNNHibatallah2023}%
  \BibitemOpen
  \bibfield  {author} {\bibinfo {author} {\bibfnamefont {M.}~\bibnamefont
  {Hibat-Allah}}, \bibinfo {author} {\bibfnamefont {R.~G.}\ \bibnamefont
  {Melko}}, \ and\ \bibinfo {author} {\bibfnamefont {J.}~\bibnamefont
  {Carrasquilla}},\ }\href@noop {} {\enquote {\bibinfo {title} {Investigating
  topological order using recurrent neural networks},}\ } (\bibinfo {year}
  {2023}),\ \Eprint {http://arxiv.org/abs/2303.11207} {arXiv:2303.11207
  [cond-mat.str-el]} \BibitemShut {NoStop}%
\bibitem [{\citenamefont {Roth}\ \emph {et~al.}(2023)\citenamefont {Roth},
  \citenamefont {Szab\'o},\ and\ \citenamefont {MacDonald}}]{AttilaGCNN2021}%
  \BibitemOpen
  \bibfield  {author} {\bibinfo {author} {\bibfnamefont {C.}~\bibnamefont
  {Roth}}, \bibinfo {author} {\bibfnamefont {A.}~\bibnamefont {Szab\'o}}, \
  and\ \bibinfo {author} {\bibfnamefont {A.~H.}\ \bibnamefont {MacDonald}},\
  }\href {\doibase 10.1103/PhysRevB.108.054410} {\bibfield  {journal} {\bibinfo
   {journal} {Phys. Rev. B}\ }\textbf {\bibinfo {volume} {108}},\ \bibinfo
  {pages} {054410} (\bibinfo {year} {2023})}\BibitemShut {NoStop}%
\bibitem [{\citenamefont {Li}\ \emph {et~al.}(2022)\citenamefont {Li},
  \citenamefont {Chen}, \citenamefont {Xiao}, \citenamefont {Wang},
  \citenamefont {Jiang}, \citenamefont {Zhao}, \citenamefont {Lin},
  \citenamefont {An}, \citenamefont {Liang},\ and\ \citenamefont
  {He}}]{LiCNN2022}%
  \BibitemOpen
  \bibfield  {author} {\bibinfo {author} {\bibfnamefont {M.}~\bibnamefont
  {Li}}, \bibinfo {author} {\bibfnamefont {J.}~\bibnamefont {Chen}}, \bibinfo
  {author} {\bibfnamefont {Q.}~\bibnamefont {Xiao}}, \bibinfo {author}
  {\bibfnamefont {F.}~\bibnamefont {Wang}}, \bibinfo {author} {\bibfnamefont
  {Q.}~\bibnamefont {Jiang}}, \bibinfo {author} {\bibfnamefont
  {X.}~\bibnamefont {Zhao}}, \bibinfo {author} {\bibfnamefont {R.}~\bibnamefont
  {Lin}}, \bibinfo {author} {\bibfnamefont {H.}~\bibnamefont {An}}, \bibinfo
  {author} {\bibfnamefont {X.}~\bibnamefont {Liang}}, \ and\ \bibinfo {author}
  {\bibfnamefont {L.}~\bibnamefont {He}},\ }\href {\doibase
  10.1109/TPDS.2022.3145163} {\bibfield  {journal} {\bibinfo  {journal} {IEEE
  Transactions on Parallel amp; Distributed Systems}\ }\textbf {\bibinfo
  {volume} {33}},\ \bibinfo {pages} {2846} (\bibinfo {year}
  {2022})}\BibitemShut {NoStop}%
\bibitem [{\citenamefont {Chen}\ and\ \citenamefont
  {Heyl}(2023)}]{HeylCNN2023}%
  \BibitemOpen
  \bibfield  {author} {\bibinfo {author} {\bibfnamefont {A.}~\bibnamefont
  {Chen}}\ and\ \bibinfo {author} {\bibfnamefont {M.}~\bibnamefont {Heyl}},\
  }\href@noop {} {\enquote {\bibinfo {title} {Efficient optimization of deep
  neural quantum states toward machine precision},}\ } (\bibinfo {year}
  {2023}),\ \Eprint {http://arxiv.org/abs/2302.01941} {arXiv:2302.01941
  [cond-mat.dis-nn]} \BibitemShut {NoStop}%
\bibitem [{\citenamefont {Liang}\ \emph {et~al.}(2022)\citenamefont {Liang},
  \citenamefont {Li}, \citenamefont {Xiao}, \citenamefont {An}, \citenamefont
  {He}, \citenamefont {Zhao}, \citenamefont {Chen}, \citenamefont {Yang},
  \citenamefont {Wang}, \citenamefont {Qian}, \citenamefont {Shen},
  \citenamefont {Jia}, \citenamefont {Gu}, \citenamefont {Liu},\ and\
  \citenamefont {Wei}}]{LiangCNN2022}%
  \BibitemOpen
  \bibfield  {author} {\bibinfo {author} {\bibfnamefont {X.}~\bibnamefont
  {Liang}}, \bibinfo {author} {\bibfnamefont {M.}~\bibnamefont {Li}}, \bibinfo
  {author} {\bibfnamefont {Q.}~\bibnamefont {Xiao}}, \bibinfo {author}
  {\bibfnamefont {H.}~\bibnamefont {An}}, \bibinfo {author} {\bibfnamefont
  {L.}~\bibnamefont {He}}, \bibinfo {author} {\bibfnamefont {X.}~\bibnamefont
  {Zhao}}, \bibinfo {author} {\bibfnamefont {J.}~\bibnamefont {Chen}}, \bibinfo
  {author} {\bibfnamefont {C.}~\bibnamefont {Yang}}, \bibinfo {author}
  {\bibfnamefont {F.}~\bibnamefont {Wang}}, \bibinfo {author} {\bibfnamefont
  {H.}~\bibnamefont {Qian}}, \bibinfo {author} {\bibfnamefont {L.}~\bibnamefont
  {Shen}}, \bibinfo {author} {\bibfnamefont {D.}~\bibnamefont {Jia}}, \bibinfo
  {author} {\bibfnamefont {Y.}~\bibnamefont {Gu}}, \bibinfo {author}
  {\bibfnamefont {X.}~\bibnamefont {Liu}}, \ and\ \bibinfo {author}
  {\bibfnamefont {Z.}~\bibnamefont {Wei}},\ }\href@noop {} {\enquote {\bibinfo
  {title} {$2^{1296}$ exponentially complex quantum many-body simulation via
  scalable deep learning method},}\ } (\bibinfo {year} {2022}),\ \Eprint
  {http://arxiv.org/abs/2204.07816} {arXiv:2204.07816 [quant-ph]} \BibitemShut
  {NoStop}%
\bibitem [{\citenamefont {Gong}\ \emph {et~al.}(2014)\citenamefont {Gong},
  \citenamefont {Zhu}, \citenamefont {Sheng}, \citenamefont {Motrunich},\ and\
  \citenamefont {Fisher}}]{DMRGSheng2014}%
  \BibitemOpen
  \bibfield  {author} {\bibinfo {author} {\bibfnamefont {S.-S.}\ \bibnamefont
  {Gong}}, \bibinfo {author} {\bibfnamefont {W.}~\bibnamefont {Zhu}}, \bibinfo
  {author} {\bibfnamefont {D.~N.}\ \bibnamefont {Sheng}}, \bibinfo {author}
  {\bibfnamefont {O.~I.}\ \bibnamefont {Motrunich}}, \ and\ \bibinfo {author}
  {\bibfnamefont {M.~P.~A.}\ \bibnamefont {Fisher}},\ }\href {\doibase
  10.1103/PhysRevLett.113.027201} {\bibfield  {journal} {\bibinfo  {journal}
  {Phys. Rev. Lett.}\ }\textbf {\bibinfo {volume} {113}},\ \bibinfo {pages}
  {027201} (\bibinfo {year} {2014})}\BibitemShut {NoStop}%
\bibitem [{\citenamefont {Ledinauskas}\ and\ \citenamefont
  {Anisimovas}(2023)}]{LedinNN2023}%
  \BibitemOpen
  \bibfield  {author} {\bibinfo {author} {\bibfnamefont {E.}~\bibnamefont
  {Ledinauskas}}\ and\ \bibinfo {author} {\bibfnamefont {E.}~\bibnamefont
  {Anisimovas}},\ }\href@noop {} {\enquote {\bibinfo {title} {Scalable
  imaginary time evolution with neural network quantum states},}\ } (\bibinfo
  {year} {2023}),\ \Eprint {http://arxiv.org/abs/2307.15521} {arXiv:2307.15521
  [quant-ph]} \BibitemShut {NoStop}%
\bibitem [{\citenamefont {Liang}\ \emph {et~al.}(2021)\citenamefont {Liang},
  \citenamefont {Dong},\ and\ \citenamefont {He}}]{LiangPEPS2021}%
  \BibitemOpen
  \bibfield  {author} {\bibinfo {author} {\bibfnamefont {X.}~\bibnamefont
  {Liang}}, \bibinfo {author} {\bibfnamefont {S.-J.}\ \bibnamefont {Dong}}, \
  and\ \bibinfo {author} {\bibfnamefont {L.}~\bibnamefont {He}},\ }\href
  {\doibase 10.1103/PhysRevB.103.035138} {\bibfield  {journal} {\bibinfo
  {journal} {Phys. Rev. B}\ }\textbf {\bibinfo {volume} {103}},\ \bibinfo
  {pages} {035138} (\bibinfo {year} {2021})}\BibitemShut {NoStop}%
\bibitem [{\citenamefont {Wang}\ \emph {et~al.}(2023)\citenamefont {Wang},
  \citenamefont {He},\ and\ \citenamefont {Lu}}]{aCNNWang2023}%
  \BibitemOpen
  \bibfield  {author} {\bibinfo {author} {\bibfnamefont {J.-Q.}\ \bibnamefont
  {Wang}}, \bibinfo {author} {\bibfnamefont {R.-Q.}\ \bibnamefont {He}}, \ and\
  \bibinfo {author} {\bibfnamefont {Z.-Y.}\ \bibnamefont {Lu}},\ }\href@noop {}
  {\enquote {\bibinfo {title} {Variational optimization of the amplitude of
  neural-network quantum many-body ground states},}\ } (\bibinfo {year}
  {2023}),\ \Eprint {http://arxiv.org/abs/2308.09664} {arXiv:2308.09664
  [cond-mat.str-el]} \BibitemShut {NoStop}%
\bibitem [{\citenamefont {Reh}\ \emph {et~al.}(2023)\citenamefont {Reh},
  \citenamefont {Schmitt},\ and\ \citenamefont {G\"arttner}}]{SchmittCNN2023}%
  \BibitemOpen
  \bibfield  {author} {\bibinfo {author} {\bibfnamefont {M.}~\bibnamefont
  {Reh}}, \bibinfo {author} {\bibfnamefont {M.}~\bibnamefont {Schmitt}}, \ and\
  \bibinfo {author} {\bibfnamefont {M.}~\bibnamefont {G\"arttner}},\ }\href
  {\doibase 10.1103/PhysRevB.107.195115} {\bibfield  {journal} {\bibinfo
  {journal} {Phys. Rev. B}\ }\textbf {\bibinfo {volume} {107}},\ \bibinfo
  {pages} {195115} (\bibinfo {year} {2023})}\BibitemShut {NoStop}%
\bibitem [{\citenamefont {Chen}\ \emph {et~al.}(2022)\citenamefont {Chen},
  \citenamefont {Hendry}, \citenamefont {Weinberg},\ and\ \citenamefont
  {Feiguin}}]{ChenRBM2022}%
  \BibitemOpen
  \bibfield  {author} {\bibinfo {author} {\bibfnamefont {H.}~\bibnamefont
  {Chen}}, \bibinfo {author} {\bibfnamefont {D.}~\bibnamefont {Hendry}},
  \bibinfo {author} {\bibfnamefont {P.}~\bibnamefont {Weinberg}}, \ and\
  \bibinfo {author} {\bibfnamefont {A.}~\bibnamefont {Feiguin}},\ }in\ \href
  {https://proceedings.neurips.cc/paper_files/paper/2022/file/3173c427cb4ed2d5eaab029c17f221ae-Paper-Conference.pdf}
  {\emph {\bibinfo {booktitle} {Advances in Neural Information Processing
  Systems}}},\ Vol.~\bibinfo {volume} {35},\ \bibinfo {editor} {edited by\
  \bibinfo {editor} {\bibfnamefont {S.}~\bibnamefont {Koyejo}}, \bibinfo
  {editor} {\bibfnamefont {S.}~\bibnamefont {Mohamed}}, \bibinfo {editor}
  {\bibfnamefont {A.}~\bibnamefont {Agarwal}}, \bibinfo {editor} {\bibfnamefont
  {D.}~\bibnamefont {Belgrave}}, \bibinfo {editor} {\bibfnamefont
  {K.}~\bibnamefont {Cho}}, \ and\ \bibinfo {editor} {\bibfnamefont
  {A.}~\bibnamefont {Oh}}}\ (\bibinfo  {publisher} {Curran Associates, Inc.},\
  \bibinfo {year} {2022})\ pp.\ \bibinfo {pages} {7490--7503}\BibitemShut
  {NoStop}%
\bibitem [{\citenamefont {Nomura}\ and\ \citenamefont
  {Imada}(2021)}]{imada-prx}%
  \BibitemOpen
  \bibfield  {author} {\bibinfo {author} {\bibfnamefont {Y.}~\bibnamefont
  {Nomura}}\ and\ \bibinfo {author} {\bibfnamefont {M.}~\bibnamefont {Imada}},\
  }\href {\doibase 10.1103/PhysRevX.11.031034} {\bibfield  {journal} {\bibinfo
  {journal} {Phys. Rev. X}\ }\textbf {\bibinfo {volume} {11}},\ \bibinfo
  {pages} {031034} (\bibinfo {year} {2021})}\BibitemShut {NoStop}%
\bibitem [{\citenamefont {Viteritti}\ \emph
  {et~al.}(2023{\natexlab{a}})\citenamefont {Viteritti}, \citenamefont
  {Rende},\ and\ \citenamefont {Becca}}]{ViT2023}%
  \BibitemOpen
  \bibfield  {author} {\bibinfo {author} {\bibfnamefont {L.~L.}\ \bibnamefont
  {Viteritti}}, \bibinfo {author} {\bibfnamefont {R.}~\bibnamefont {Rende}}, \
  and\ \bibinfo {author} {\bibfnamefont {F.}~\bibnamefont {Becca}},\ }\href
  {\doibase 10.1103/PhysRevLett.130.236401} {\bibfield  {journal} {\bibinfo
  {journal} {Phys. Rev. Lett.}\ }\textbf {\bibinfo {volume} {130}},\ \bibinfo
  {pages} {236401} (\bibinfo {year} {2023}{\natexlab{a}})}\BibitemShut
  {NoStop}%
\bibitem [{\citenamefont {Viteritti}\ \emph
  {et~al.}(2023{\natexlab{b}})\citenamefont {Viteritti}, \citenamefont {Rende},
  \citenamefont {Parola}, \citenamefont {Goldt},\ and\ \citenamefont
  {Becca}}]{shastry2023}%
  \BibitemOpen
  \bibfield  {author} {\bibinfo {author} {\bibfnamefont {L.~L.}\ \bibnamefont
  {Viteritti}}, \bibinfo {author} {\bibfnamefont {R.}~\bibnamefont {Rende}},
  \bibinfo {author} {\bibfnamefont {A.}~\bibnamefont {Parola}}, \bibinfo
  {author} {\bibfnamefont {S.}~\bibnamefont {Goldt}}, \ and\ \bibinfo {author}
  {\bibfnamefont {F.}~\bibnamefont {Becca}},\ }\href@noop {} {\enquote
  {\bibinfo {title} {Transformer wave function for the shastry-sutherland
  model: emergence of a spin-liquid phase},}\ } (\bibinfo {year}
  {2023}{\natexlab{b}}),\ \Eprint {http://arxiv.org/abs/2311.16889}
  {arXiv:2311.16889 [cond-mat.str-el]} \BibitemShut {NoStop}%
\bibitem [{\citenamefont {Sprague}\ and\ \citenamefont
  {Czischek}(2023)}]{Czischek2023}%
  \BibitemOpen
  \bibfield  {author} {\bibinfo {author} {\bibfnamefont {K.}~\bibnamefont
  {Sprague}}\ and\ \bibinfo {author} {\bibfnamefont {S.}~\bibnamefont
  {Czischek}},\ }\href@noop {} {\enquote {\bibinfo {title} {Variational monte
  carlo with large patched transformers},}\ } (\bibinfo {year} {2023}),\
  \Eprint {http://arxiv.org/abs/2306.03921} {arXiv:2306.03921 [quant-ph]}
  \BibitemShut {NoStop}%
\bibitem [{\citenamefont {Luo}\ \emph {et~al.}(2023)\citenamefont {Luo},
  \citenamefont {Chen}, \citenamefont {Hu}, \citenamefont {Zhao}, \citenamefont
  {Hur},\ and\ \citenamefont {Clark}}]{DiLuo2023}%
  \BibitemOpen
  \bibfield  {author} {\bibinfo {author} {\bibfnamefont {D.}~\bibnamefont
  {Luo}}, \bibinfo {author} {\bibfnamefont {Z.}~\bibnamefont {Chen}}, \bibinfo
  {author} {\bibfnamefont {K.}~\bibnamefont {Hu}}, \bibinfo {author}
  {\bibfnamefont {Z.}~\bibnamefont {Zhao}}, \bibinfo {author} {\bibfnamefont
  {V.~M.}\ \bibnamefont {Hur}}, \ and\ \bibinfo {author} {\bibfnamefont
  {B.~K.}\ \bibnamefont {Clark}},\ }\href {\doibase
  10.1103/PhysRevResearch.5.013216} {\bibfield  {journal} {\bibinfo  {journal}
  {Phys. Rev. Res.}\ }\textbf {\bibinfo {volume} {5}},\ \bibinfo {pages}
  {013216} (\bibinfo {year} {2023})}\BibitemShut {NoStop}%
\bibitem [{\citenamefont {Marshall}(1955)}]{marshall1955}%
  \BibitemOpen
  \bibfield  {author} {\bibinfo {author} {\bibfnamefont {W.}~\bibnamefont
  {Marshall}},\ }\href {http://www.jstor.org/stable/99682} {\bibfield
  {journal} {\bibinfo  {journal} {Proceedings of the Royal Society of London.
  Series A, Mathematical and Physical Sciences}\ }\textbf {\bibinfo {volume}
  {232}},\ \bibinfo {pages} {48} (\bibinfo {year} {1955})}\BibitemShut
  {NoStop}%
\bibitem [{\citenamefont {Becca}\ and\ \citenamefont
  {Sorella}(2017)}]{BeccaSorella2017}%
  \BibitemOpen
  \bibfield  {author} {\bibinfo {author} {\bibfnamefont {F.}~\bibnamefont
  {Becca}}\ and\ \bibinfo {author} {\bibfnamefont {S.}~\bibnamefont
  {Sorella}},\ }\href {\doibase 10.1017/9781316417041} {\emph {\bibinfo {title}
  {Quantum Monte Carlo Approaches for Correlated Systems}}}\ (\bibinfo
  {publisher} {Cambridge University Press},\ \bibinfo {year}
  {2017})\BibitemShut {NoStop}%
\bibitem [{\citenamefont {Henderson}\ and\ \citenamefont
  {Searle}(1981)}]{henderson1981}%
  \BibitemOpen
  \bibfield  {author} {\bibinfo {author} {\bibfnamefont {H.~V.}\ \bibnamefont
  {Henderson}}\ and\ \bibinfo {author} {\bibfnamefont {S.~R.}\ \bibnamefont
  {Searle}},\ }\href {https://api.semanticscholar.org/CorpusID:123113511}
  {\bibfield  {journal} {\bibinfo  {journal} {Siam Review}\ }\textbf {\bibinfo
  {volume} {23}},\ \bibinfo {pages} {53} (\bibinfo {year} {1981})}\BibitemShut
  {NoStop}%
\bibitem [{\citenamefont {Petersen}\ and\ \citenamefont
  {Pedersen}(2012)}]{matrix-cookbook}%
  \BibitemOpen
  \bibfield  {author} {\bibinfo {author} {\bibfnamefont {K.~B.}\ \bibnamefont
  {Petersen}}\ and\ \bibinfo {author} {\bibfnamefont {M.~S.}\ \bibnamefont
  {Pedersen}},\ }\href {http://www2.compute.dtu.dk/pubdb/pubs/3274-full.html}
  {\enquote {\bibinfo {title} {The matrix cookbook},}\ } (\bibinfo {year}
  {2012})\BibitemShut {NoStop}%
\bibitem [{\citenamefont {Novak}\ \emph {et~al.}(2022)\citenamefont {Novak},
  \citenamefont {Sohl-Dickstein},\ and\ \citenamefont {Schoenholz}}]{novak22}%
  \BibitemOpen
  \bibfield  {author} {\bibinfo {author} {\bibfnamefont {R.}~\bibnamefont
  {Novak}}, \bibinfo {author} {\bibfnamefont {J.}~\bibnamefont
  {Sohl-Dickstein}}, \ and\ \bibinfo {author} {\bibfnamefont {S.~S.}\
  \bibnamefont {Schoenholz}},\ }in\ \href
  {https://proceedings.mlr.press/v162/novak22a.html} {\emph {\bibinfo
  {booktitle} {Proceedings of the 39th International Conference on Machine
  Learning}}},\ \bibinfo {series} {Proceedings of Machine Learning Research},
  Vol.\ \bibinfo {volume} {162}\ (\bibinfo  {publisher} {PMLR},\ \bibinfo
  {year} {2022})\ pp.\ \bibinfo {pages} {17018--17044}\BibitemShut {NoStop}%
\bibitem [{\citenamefont {Vicentini}\ \emph {et~al.}(2022)\citenamefont
  {Vicentini}, \citenamefont {Hofmann}, \citenamefont {Szabó}, \citenamefont
  {Wu}, \citenamefont {Roth}, \citenamefont {Giuliani}, \citenamefont {Pescia},
  \citenamefont {Nys}, \citenamefont {Vargas-Calderón}, \citenamefont
  {Astrakhantsev},\ and\ \citenamefont {Carleo}}]{netket3}%
  \BibitemOpen
  \bibfield  {author} {\bibinfo {author} {\bibfnamefont {F.}~\bibnamefont
  {Vicentini}}, \bibinfo {author} {\bibfnamefont {D.}~\bibnamefont {Hofmann}},
  \bibinfo {author} {\bibfnamefont {A.}~\bibnamefont {Szabó}}, \bibinfo
  {author} {\bibfnamefont {D.}~\bibnamefont {Wu}}, \bibinfo {author}
  {\bibfnamefont {C.}~\bibnamefont {Roth}}, \bibinfo {author} {\bibfnamefont
  {C.}~\bibnamefont {Giuliani}}, \bibinfo {author} {\bibfnamefont
  {G.}~\bibnamefont {Pescia}}, \bibinfo {author} {\bibfnamefont
  {J.}~\bibnamefont {Nys}}, \bibinfo {author} {\bibfnamefont {V.}~\bibnamefont
  {Vargas-Calderón}}, \bibinfo {author} {\bibfnamefont {N.}~\bibnamefont
  {Astrakhantsev}}, \ and\ \bibinfo {author} {\bibfnamefont {G.}~\bibnamefont
  {Carleo}},\ }\href {\doibase 10.21468/SciPostPhysCodeb.7} {\bibfield
  {journal} {\bibinfo  {journal} {SciPost Phys. Codebases}\ ,\ \bibinfo {pages}
  {7}} (\bibinfo {year} {2022})}\BibitemShut {NoStop}%
\bibitem [{\citenamefont {Lovato}\ \emph {et~al.}(2022)\citenamefont {Lovato},
  \citenamefont {Adams}, \citenamefont {Carleo},\ and\ \citenamefont
  {Rocco}}]{lovato2022}%
  \BibitemOpen
  \bibfield  {author} {\bibinfo {author} {\bibfnamefont {A.}~\bibnamefont
  {Lovato}}, \bibinfo {author} {\bibfnamefont {C.}~\bibnamefont {Adams}},
  \bibinfo {author} {\bibfnamefont {G.}~\bibnamefont {Carleo}}, \ and\ \bibinfo
  {author} {\bibfnamefont {N.}~\bibnamefont {Rocco}},\ }\href {\doibase
  10.1103/PhysRevResearch.4.043178} {\bibfield  {journal} {\bibinfo  {journal}
  {Phys. Rev. Res.}\ }\textbf {\bibinfo {volume} {4}},\ \bibinfo {pages}
  {043178} (\bibinfo {year} {2022})}\BibitemShut {NoStop}%
\bibitem [{\citenamefont {Rende}\ \emph {et~al.}(2023)\citenamefont {Rende},
  \citenamefont {Gerace}, \citenamefont {Laio},\ and\ \citenamefont
  {Goldt}}]{rende2023optimal}%
  \BibitemOpen
  \bibfield  {author} {\bibinfo {author} {\bibfnamefont {R.}~\bibnamefont
  {Rende}}, \bibinfo {author} {\bibfnamefont {F.}~\bibnamefont {Gerace}},
  \bibinfo {author} {\bibfnamefont {A.}~\bibnamefont {Laio}}, \ and\ \bibinfo
  {author} {\bibfnamefont {S.}~\bibnamefont {Goldt}},\ }\href@noop {} {\enquote
  {\bibinfo {title} {Optimal inference of a generalised potts model by
  single-layer transformers with factored attention},}\ } (\bibinfo {year}
  {2023}),\ \Eprint {http://arxiv.org/abs/2304.07235} {arXiv:2304.07235}
  \BibitemShut {NoStop}%
\bibitem [{\citenamefont {Bhattacharya}\ \emph {et~al.}(2020)\citenamefont
  {Bhattacharya}, \citenamefont {Thomas}, \citenamefont {Rao}, \citenamefont
  {Daupras}, \citenamefont {Koo}, \citenamefont {Baker}, \citenamefont {Song},\
  and\ \citenamefont {Ovchinnikov}}]{bhattacharya2020}%
  \BibitemOpen
  \bibfield  {author} {\bibinfo {author} {\bibfnamefont {N.}~\bibnamefont
  {Bhattacharya}}, \bibinfo {author} {\bibfnamefont {N.}~\bibnamefont
  {Thomas}}, \bibinfo {author} {\bibfnamefont {R.}~\bibnamefont {Rao}},
  \bibinfo {author} {\bibfnamefont {J.}~\bibnamefont {Daupras}}, \bibinfo
  {author} {\bibfnamefont {P.}~\bibnamefont {Koo}}, \bibinfo {author}
  {\bibfnamefont {D.}~\bibnamefont {Baker}}, \bibinfo {author} {\bibfnamefont
  {Y.}~\bibnamefont {Song}}, \ and\ \bibinfo {author} {\bibfnamefont
  {S.}~\bibnamefont {Ovchinnikov}},\ }\href {\doibase
  10.1101/2020.12.21.423882} {\enquote {\bibinfo {title} {Single layers of
  attention suffice to predict protein contacts},}\ } (\bibinfo {year}
  {2020})\BibitemShut {NoStop}%
\bibitem [{\citenamefont {Rende}\ and\ \citenamefont
  {Viteritti}(2024)}]{rende2024queries}%
  \BibitemOpen
  \bibfield  {author} {\bibinfo {author} {\bibfnamefont {R.}~\bibnamefont
  {Rende}}\ and\ \bibinfo {author} {\bibfnamefont {L.~L.}\ \bibnamefont
  {Viteritti}},\ }\href@noop {} {\enquote {\bibinfo {title} {Are queries and
  keys always relevant? a case study on transformer wave functions},}\ }
  (\bibinfo {year} {2024}),\ \Eprint {http://arxiv.org/abs/2405.18874}
  {arXiv:2405.18874 [cond-mat.dis-nn]} \BibitemShut {NoStop}%
\bibitem [{\citenamefont {Dosovitskiy}\ \emph {et~al.}(2021)\citenamefont
  {Dosovitskiy}, \citenamefont {Beyer}, \citenamefont {Kolesnikov},
  \citenamefont {Weissenborn}, \citenamefont {Zhai}, \citenamefont
  {Unterthiner}, \citenamefont {Dehghani}, \citenamefont {Minderer},
  \citenamefont {Heigold}, \citenamefont {Gelly}, \citenamefont {Uszkoreit},\
  and\ \citenamefont {Houlsby}}]{dosovitskiy2021image}%
  \BibitemOpen
  \bibfield  {author} {\bibinfo {author} {\bibfnamefont {A.}~\bibnamefont
  {Dosovitskiy}}, \bibinfo {author} {\bibfnamefont {L.}~\bibnamefont {Beyer}},
  \bibinfo {author} {\bibfnamefont {A.}~\bibnamefont {Kolesnikov}}, \bibinfo
  {author} {\bibfnamefont {D.}~\bibnamefont {Weissenborn}}, \bibinfo {author}
  {\bibfnamefont {X.}~\bibnamefont {Zhai}}, \bibinfo {author} {\bibfnamefont
  {T.}~\bibnamefont {Unterthiner}}, \bibinfo {author} {\bibfnamefont
  {M.}~\bibnamefont {Dehghani}}, \bibinfo {author} {\bibfnamefont
  {M.}~\bibnamefont {Minderer}}, \bibinfo {author} {\bibfnamefont
  {G.}~\bibnamefont {Heigold}}, \bibinfo {author} {\bibfnamefont
  {S.}~\bibnamefont {Gelly}}, \bibinfo {author} {\bibfnamefont
  {J.}~\bibnamefont {Uszkoreit}}, \ and\ \bibinfo {author} {\bibfnamefont
  {N.}~\bibnamefont {Houlsby}},\ }\href@noop {} {\enquote {\bibinfo {title} {An
  image is worth 16x16 words: Transformers for image recognition at scale},}\ }
  (\bibinfo {year} {2021}),\ \Eprint {http://arxiv.org/abs/2010.11929}
  {arXiv:2010.11929 [cs.CV]} \BibitemShut {NoStop}%
\bibitem [{\citenamefont {Xiong}\ \emph {et~al.}(2020)\citenamefont {Xiong},
  \citenamefont {Yang}, \citenamefont {He}, \citenamefont {Zheng},
  \citenamefont {Zheng}, \citenamefont {Xing}, \citenamefont {Zhang},
  \citenamefont {Lan}, \citenamefont {Wang},\ and\ \citenamefont
  {Liu}}]{xiong2020layer}%
  \BibitemOpen
  \bibfield  {author} {\bibinfo {author} {\bibfnamefont {R.}~\bibnamefont
  {Xiong}}, \bibinfo {author} {\bibfnamefont {Y.}~\bibnamefont {Yang}},
  \bibinfo {author} {\bibfnamefont {D.}~\bibnamefont {He}}, \bibinfo {author}
  {\bibfnamefont {K.}~\bibnamefont {Zheng}}, \bibinfo {author} {\bibfnamefont
  {S.}~\bibnamefont {Zheng}}, \bibinfo {author} {\bibfnamefont
  {C.}~\bibnamefont {Xing}}, \bibinfo {author} {\bibfnamefont {H.}~\bibnamefont
  {Zhang}}, \bibinfo {author} {\bibfnamefont {Y.}~\bibnamefont {Lan}}, \bibinfo
  {author} {\bibfnamefont {L.}~\bibnamefont {Wang}}, \ and\ \bibinfo {author}
  {\bibfnamefont {T.-Y.}\ \bibnamefont {Liu}},\ }\href@noop {} {\enquote
  {\bibinfo {title} {On layer normalization in the transformer architecture},}\
  } (\bibinfo {year} {2020}),\ \Eprint {http://arxiv.org/abs/2002.04745}
  {arXiv:2002.04745 [cs.LG]} \BibitemShut {NoStop}%
\bibitem [{\citenamefont {Nomura}(2021)}]{Nomura_2021}%
  \BibitemOpen
  \bibfield  {author} {\bibinfo {author} {\bibfnamefont {Y.}~\bibnamefont
  {Nomura}},\ }\href {\doibase 10.1088/1361-648X/abe268} {\bibfield  {journal}
  {\bibinfo  {journal} {Journal of Physics: Condensed Matter}\ }\textbf
  {\bibinfo {volume} {33}},\ \bibinfo {pages} {174003} (\bibinfo {year}
  {2021})}\BibitemShut {NoStop}%
\bibitem [{\citenamefont {Urban}\ \emph {et~al.}(2017)\citenamefont {Urban},
  \citenamefont {Geras}, \citenamefont {Kahou}, \citenamefont {Aslan},
  \citenamefont {Wang}, \citenamefont {Mohamed}, \citenamefont {Philipose},
  \citenamefont {Richardson},\ and\ \citenamefont {Caruana}}]{urban2017do}%
  \BibitemOpen
  \bibfield  {author} {\bibinfo {author} {\bibfnamefont {G.}~\bibnamefont
  {Urban}}, \bibinfo {author} {\bibfnamefont {K.~J.}\ \bibnamefont {Geras}},
  \bibinfo {author} {\bibfnamefont {S.~E.}\ \bibnamefont {Kahou}}, \bibinfo
  {author} {\bibfnamefont {O.}~\bibnamefont {Aslan}}, \bibinfo {author}
  {\bibfnamefont {S.}~\bibnamefont {Wang}}, \bibinfo {author} {\bibfnamefont
  {A.}~\bibnamefont {Mohamed}}, \bibinfo {author} {\bibfnamefont
  {M.}~\bibnamefont {Philipose}}, \bibinfo {author} {\bibfnamefont
  {M.}~\bibnamefont {Richardson}}, \ and\ \bibinfo {author} {\bibfnamefont
  {R.}~\bibnamefont {Caruana}},\ }in\ \href
  {https://openreview.net/forum?id=r10FA8Kxg} {\emph {\bibinfo {booktitle}
  {International Conference on Learning Representations}}}\ (\bibinfo {year}
  {2017})\BibitemShut {NoStop}%
\bibitem [{\citenamefont {d\textquotesingle Ascoli}\ \emph
  {et~al.}(2019)\citenamefont {d\textquotesingle Ascoli}, \citenamefont
  {Sagun}, \citenamefont {Biroli},\ and\ \citenamefont {Bruna}}]{Ascoli2019}%
  \BibitemOpen
  \bibfield  {author} {\bibinfo {author} {\bibfnamefont {S.}~\bibnamefont
  {d\textquotesingle Ascoli}}, \bibinfo {author} {\bibfnamefont
  {L.}~\bibnamefont {Sagun}}, \bibinfo {author} {\bibfnamefont
  {G.}~\bibnamefont {Biroli}}, \ and\ \bibinfo {author} {\bibfnamefont
  {J.}~\bibnamefont {Bruna}},\ }in\ \href
  {https://proceedings.neurips.cc/paper_files/paper/2019/file/124c3e4ada4a529aa0fedece80bb42ab-Paper.pdf}
  {\emph {\bibinfo {booktitle} {Advances in Neural Information Processing
  Systems}}},\ Vol.~\bibinfo {volume} {32},\ \bibinfo {editor} {edited by\
  \bibinfo {editor} {\bibfnamefont {H.}~\bibnamefont {Wallach}}, \bibinfo
  {editor} {\bibfnamefont {H.}~\bibnamefont {Larochelle}}, \bibinfo {editor}
  {\bibfnamefont {A.}~\bibnamefont {Beygelzimer}}, \bibinfo {editor}
  {\bibfnamefont {F.}~\bibnamefont {d\textquotesingle Alch\'{e}-Buc}}, \bibinfo
  {editor} {\bibfnamefont {E.}~\bibnamefont {Fox}}, \ and\ \bibinfo {editor}
  {\bibfnamefont {R.}~\bibnamefont {Garnett}}}\ (\bibinfo  {publisher} {Curran
  Associates, Inc.},\ \bibinfo {year} {2019})\BibitemShut {NoStop}%
\bibitem [{\citenamefont {Ingrosso}\ and\ \citenamefont
  {Goldt}(2022)}]{ingrosso2022}%
  \BibitemOpen
  \bibfield  {author} {\bibinfo {author} {\bibfnamefont {A.}~\bibnamefont
  {Ingrosso}}\ and\ \bibinfo {author} {\bibfnamefont {S.}~\bibnamefont
  {Goldt}},\ }\href {\doibase 10.1073/pnas.2201854119} {\bibfield  {journal}
  {\bibinfo  {journal} {Proceedings of the National Academy of Sciences}\
  }\textbf {\bibinfo {volume} {119}},\ \bibinfo {pages} {e2201854119} (\bibinfo
  {year} {2022})},\ \Eprint
  {http://arxiv.org/abs/https://www.pnas.org/doi/pdf/10.1073/pnas.2201854119}
  {https://www.pnas.org/doi/pdf/10.1073/pnas.2201854119} \BibitemShut {NoStop}%
\bibitem [{\citenamefont {Mendes-Santos}\ \emph {et~al.}(2023)\citenamefont
  {Mendes-Santos}, \citenamefont {Schmitt},\ and\ \citenamefont
  {Heyl}}]{HeylDyn2023}%
  \BibitemOpen
  \bibfield  {author} {\bibinfo {author} {\bibfnamefont {T.}~\bibnamefont
  {Mendes-Santos}}, \bibinfo {author} {\bibfnamefont {M.}~\bibnamefont
  {Schmitt}}, \ and\ \bibinfo {author} {\bibfnamefont {M.}~\bibnamefont
  {Heyl}},\ }\href {\doibase 10.1103/PhysRevLett.131.046501} {\bibfield
  {journal} {\bibinfo  {journal} {Phys. Rev. Lett.}\ }\textbf {\bibinfo
  {volume} {131}},\ \bibinfo {pages} {046501} (\bibinfo {year}
  {2023})}\BibitemShut {NoStop}%
\bibitem [{\citenamefont {Schmitt}\ and\ \citenamefont
  {Heyl}(2020)}]{SchmittDyn2020}%
  \BibitemOpen
  \bibfield  {author} {\bibinfo {author} {\bibfnamefont {M.}~\bibnamefont
  {Schmitt}}\ and\ \bibinfo {author} {\bibfnamefont {M.}~\bibnamefont {Heyl}},\
  }\href {\doibase 10.1103/PhysRevLett.125.100503} {\bibfield  {journal}
  {\bibinfo  {journal} {Phys. Rev. Lett.}\ }\textbf {\bibinfo {volume} {125}},\
  \bibinfo {pages} {100503} (\bibinfo {year} {2020})}\BibitemShut {NoStop}%
\bibitem [{\citenamefont {Nakano}\ \emph {et~al.}(2020)\citenamefont {Nakano},
  \citenamefont {Attaccalite}, \citenamefont {Barborini}, \citenamefont
  {Capriotti}, \citenamefont {Casula}, \citenamefont {Coccia}, \citenamefont
  {Dagrada}, \citenamefont {Genovese}, \citenamefont {Luo}, \citenamefont
  {Mazzola}, \citenamefont {Zen},\ and\ \citenamefont {Sorella}}]{Nakano_2020}%
  \BibitemOpen
  \bibfield  {author} {\bibinfo {author} {\bibfnamefont {K.}~\bibnamefont
  {Nakano}}, \bibinfo {author} {\bibfnamefont {C.}~\bibnamefont {Attaccalite}},
  \bibinfo {author} {\bibfnamefont {M.}~\bibnamefont {Barborini}}, \bibinfo
  {author} {\bibfnamefont {L.}~\bibnamefont {Capriotti}}, \bibinfo {author}
  {\bibfnamefont {M.}~\bibnamefont {Casula}}, \bibinfo {author} {\bibfnamefont
  {E.}~\bibnamefont {Coccia}}, \bibinfo {author} {\bibfnamefont
  {M.}~\bibnamefont {Dagrada}}, \bibinfo {author} {\bibfnamefont
  {C.}~\bibnamefont {Genovese}}, \bibinfo {author} {\bibfnamefont
  {Y.}~\bibnamefont {Luo}}, \bibinfo {author} {\bibfnamefont {G.}~\bibnamefont
  {Mazzola}}, \bibinfo {author} {\bibfnamefont {A.}~\bibnamefont {Zen}}, \ and\
  \bibinfo {author} {\bibfnamefont {S.}~\bibnamefont {Sorella}},\ }\href
  {\doibase 10.1063/5.0005037} {\bibfield  {journal} {\bibinfo  {journal} {The
  Journal of Chemical Physics}\ }\textbf {\bibinfo {volume} {152}} (\bibinfo
  {year} {2020}),\ 10.1063/5.0005037}\BibitemShut {NoStop}%
\bibitem [{\citenamefont {Bishop}(2006)}]{bishop2006}%
  \BibitemOpen
  \bibfield  {author} {\bibinfo {author} {\bibfnamefont {C.~M.}\ \bibnamefont
  {Bishop}},\ }\href@noop {} {\emph {\bibinfo {title} {Pattern Recognition and
  Machine Learning (Information Science and Statistics)}}}\ (\bibinfo
  {publisher} {Springer-Verlag},\ \bibinfo {address} {Berlin, Heidelberg},\
  \bibinfo {year} {2006})\BibitemShut {NoStop}%
\bibitem [{\citenamefont {Giuliani}\ \emph {et~al.}(2023)\citenamefont
  {Giuliani}, \citenamefont {Vicentini}, \citenamefont {Rossi},\ and\
  \citenamefont {Carleo}}]{giuliani2023}%
  \BibitemOpen
  \bibfield  {author} {\bibinfo {author} {\bibfnamefont {C.}~\bibnamefont
  {Giuliani}}, \bibinfo {author} {\bibfnamefont {F.}~\bibnamefont {Vicentini}},
  \bibinfo {author} {\bibfnamefont {R.}~\bibnamefont {Rossi}}, \ and\ \bibinfo
  {author} {\bibfnamefont {G.}~\bibnamefont {Carleo}},\ }\href {\doibase
  10.22331/q-2023-08-29-1096} {\bibfield  {journal} {\bibinfo  {journal}
  {{Quantum}}\ }\textbf {\bibinfo {volume} {7}},\ \bibinfo {pages} {1096}
  (\bibinfo {year} {2023})}\BibitemShut {NoStop}%
\bibitem [{\citenamefont {Bradbury}\ \emph {et~al.}(2018)\citenamefont
  {Bradbury}, \citenamefont {Frostig}, \citenamefont {Hawkins}, \citenamefont
  {Johnson}, \citenamefont {Leary}, \citenamefont {Maclaurin}, \citenamefont
  {Necula}, \citenamefont {Paszke}, \citenamefont {Vander{P}las}, \citenamefont
  {Wanderman-{M}ilne},\ and\ \citenamefont {Zhang}}]{jax2018github}%
  \BibitemOpen
  \bibfield  {author} {\bibinfo {author} {\bibfnamefont {J.}~\bibnamefont
  {Bradbury}}, \bibinfo {author} {\bibfnamefont {R.}~\bibnamefont {Frostig}},
  \bibinfo {author} {\bibfnamefont {P.}~\bibnamefont {Hawkins}}, \bibinfo
  {author} {\bibfnamefont {M.}~\bibnamefont {Johnson}}, \bibinfo {author}
  {\bibfnamefont {C.}~\bibnamefont {Leary}}, \bibinfo {author} {\bibfnamefont
  {D.}~\bibnamefont {Maclaurin}}, \bibinfo {author} {\bibfnamefont
  {G.}~\bibnamefont {Necula}}, \bibinfo {author} {\bibfnamefont
  {A.}~\bibnamefont {Paszke}}, \bibinfo {author} {\bibfnamefont
  {J.}~\bibnamefont {Vander{P}las}}, \bibinfo {author} {\bibfnamefont
  {S.}~\bibnamefont {Wanderman-{M}ilne}}, \ and\ \bibinfo {author}
  {\bibfnamefont {Q.}~\bibnamefont {Zhang}},\ }\href
  {http://github.com/google/jax} {\enquote {\bibinfo {title} {{JAX}: composable
  transformations of {P}ython+{N}um{P}y programs},}\ } (\bibinfo {year}
  {2018})\BibitemShut {NoStop}%
\bibitem [{\citenamefont {Häfner}\ and\ \citenamefont
  {Vicentini}(2021)}]{mpi4jax}%
  \BibitemOpen
  \bibfield  {author} {\bibinfo {author} {\bibfnamefont {D.}~\bibnamefont
  {Häfner}}\ and\ \bibinfo {author} {\bibfnamefont {F.}~\bibnamefont
  {Vicentini}},\ }\href {\doibase 10.21105/joss.03419} {\bibfield  {journal}
  {\bibinfo  {journal} {Journal of Open Source Software}\ }\textbf {\bibinfo
  {volume} {6}},\ \bibinfo {pages} {3419} (\bibinfo {year} {2021})}\BibitemShut
  {NoStop}%
\end{thebibliography}%

\end{document}